\documentclass[12pt]{article}
\usepackage{amsfonts}
\usepackage{amssymb}
\usepackage{mathrsfs}
\usepackage{subfigure}
\usepackage{amsmath,amsthm,amssymb,amscd}
\usepackage[all]{xy}
\usepackage{hyperref}
\usepackage{multirow}
\usepackage{color}
\usepackage{amsfonts}
\usepackage{float}
\usepackage{mathtools,slashed}
\usepackage{bbold}
\usepackage{amsmath}
\usepackage{bbm}
\usepackage{slashed}
\usepackage[utf8]{inputenc}
\usepackage{color}
\usepackage[ruled,vlined]{algorithm2e}

\usepackage{cite}

\def\eqa{\begin{eqnarray}}
\def\eqae{\end{eqnarray}}
\def\eq{\begin{equation}}
\def\eqe{\end{equation}}
\def\be{\begin{equation}}
\def\ee{\end{equation}}
\def\bea{\begin{eqnarray}}
\def\eea{\end{eqnarray}}
\def\ba{\begin{array}}
\def\ea{\end{array}}
\def\bd{\begin{displaymath}}
\def\ed{\end{displaymath}}

\def\>{\rangle}
\def\<{\langle}
\def\a{\alpha}

\def\th{\theta}

\newcommand{\fft}[2]{\frac{#1}{#2}}

\numberwithin{equation}{section}

\newcommand{\tx}{\tilde{x}}

\newcommand{\tabincell}[2]{\begin{tabular}{@{}#1@{}}#2\end{tabular}}  

\begin{document}

\begin{titlepage}
\hfill LCTP-19-16

\vskip 1 cm

\begin{center}
{\large \bf The Topologically Twisted Index  in the 't Hooft Limit}\\

\vskip .7cm

{\large \bf  and the Dual AdS$_4$ Black Hole Entropy}\\

\end{center}

\vskip .7 cm

\vskip 1 cm
\begin{center}
{ \large Leopoldo A. Pando Zayas${}^{a,b}$ and Yu Xin${}^a$}
\end{center}

\vskip .4cm \centerline{\it ${}^a$ Leinweber Center for Theoretical
Physics,   Randall Laboratory of Physics}
\centerline{ \it The University of
Michigan, Ann Arbor, MI 48109, USA}
\bigskip\bigskip

\centerline{\it ${}^b$ The Abdus Salam International Centre for Theoretical Physics}
\centerline{\it  Strada Costiera 11,  34014 Trieste, Italy}

\bigskip\bigskip

\vskip 1 cm

\vskip 1.5 cm
\begin{abstract}

We study the topologically twisted index of $\mathcal{N}=6$ supersymmetric Chern-Simons matter theory with $U(N)_k \times U(N)_{-k}$ gauge group  in the 't Hooft limit, that is, for $N, k \, \to \infty$ with $\lambda=N/k$ fixed.  In the regime where $\lambda$ is fixed and large we find an analytical expression for the leading order term of the index. The leading term of the index matches precisely the Bekenstein-Hawking entropy of the dual asymptotically AdS$_4$ magnetically charged black holes embedded in IIA supergrvity on AdS$_4\times \mathbb{CP}^3$, after a standard Legendre transformation.  We numerically explore the genus expansion of the topologically twisteed index beyond the leading order, focusing on the genus one contribution, that is, $N^0$. We find qualitative agreement with the topological expansion of the free energy on $S^3$ at genus one. Our logarithmic in $\lambda$ term constitutes a prediction for the one-loop effective action on the IIA supergravity side. 
\end{abstract}

\vskip  1.5 cm

{\tt lpandoz@umich.edu,  yxinee@umich.edu}
\end{titlepage}

\tableofcontents

\section{Introduction}

One of the best established pairs of the AdS/CFT correspondence is the duality between Chern-Simons matter  theory with gauge group $U(N)_{k}\times U(N)_{-k}$ and string theory. In the large $N$ limit there are, in fact,  two distinctive regimes: for large $N$ and fixed $k$ the dual gravity is found in M-theory while in the 't Hooft limit, that is, large $N$ and $k$ with $\lambda=N/k$ fixed, the gravity theory is low energy string theory on AdS$_4\times \mathbb{CP}^3$  \cite{Aharony:2008ug}.

The difference between the M-theory limit and the IIA limit has been one of the most interesting aspects of this duality since its very early inception. In particular the $N^{3/2}$ versus the $N^2$ growth in the number of the degrees of freedom has been one of the most enigmatic problems that the correspondence has elucidated \cite{Drukker:2010nc}. The clarification was made manifest by studying the free energy on $S^3$ which cleanly captures the  two behaviors  \cite{Drukker:2010nc}; such analysis has also shed light on various non-perturbative aspects \cite{Drukker:2011zy}.  It is thus expected that other observables related to counting of degrees of freedom should follow and enrich this dichotomy. One such observable is the topologically twisted index   \cite{Benini:2015noa,Honda:2015yha,Closset:2015rna,Benini:2016hjo,Closset:2016arn,Closset:2017zgf,Closset:2018ghr}. 

There has recently been a remarkable development in our understanding of the microscopic origin of the entropy for asymptotically AdS$_4$ black holes precisely via studies of the topologically twisted index  \cite{Benini:2015eyy,Benini:2016rke,Cabo-Bizet:2017jsl,Azzurli:2017kxo, Hosseini:2017fjo,Benini:2017oxt}, for recent reviews of these developments, including full  lists of references, see \cite{Hosseini:2018qsx,Zaffaroni:2019dhb}.  Most of the discussion thus far has been focused in the M-theory regime where the leading term is of the order $N^{3/2}$ or $N^3$; in this manuscript we focus on the IIA regime with $N^2$ growth.  Explorations beyond the  leading order have also proven fruitful. In particular, the sub-leading logarithmic in $N$ corrections for the ABJM theory and the Chern-Simons matter theory dual to massive IIA have been discussed in \cite{Liu:2017vll} and\cite{Liu:2018bac} , respectively.  The more subtle issue of matching with the one-loop quantum supergravity has been achieved in \cite{Liu:2017vbl} after various discussions  \cite{Liu:2017vll,Jeon:2017aif} (see also \cite{Hristov:2018lod}).  More recently, sub-leading matching in the context of asymptotically AdS$_4$ black holes obtained from M5 branes wrapping  hyperbolic three-manifolds was demonstrated in \cite{Gang:2018hjd,Gang:2019uay}. The 3d-3d correspondence allows to evaluate the large $N$ partition functions using Chern-Simons topological invariants, thus providing analytic control over the results which is not the case in the theories related to M2 branes. 

In this manuscript we extend the analysis of the topologically twisted index of ABJM theory to the 't Hooft limit: $N, k \to \infty$  holding $\lambda=N/k$ fixed.  We then compare with the supergravity Bekenstein-Hawking entropy  which is  valid for large values of $\lambda$.  The leading in $N$ term for the topologically twisted index equals 
\be
{\rm Re}\log Z=-\frac{1}{3}\,\frac{N^2}{\sqrt{\lambda}}\sqrt{2\,\,\Delta_1 \Delta_2 \Delta_3\Delta_4}\sum\limits_a\frac{n_a}{\Delta_a}.
\ee
The above expression precisely reproduces, after a Legendre transformation, the entropy of the dual magnetically charged black holes asymptotically to AdS$_4\times \mathbb{CP}^3$.  We emphasize that the matching takes place for large values of $\lambda$ and further demonstrate numerically that the fitting improves for large values of $\lambda$.

We also study the first sub-leading in $N$ term, namely, the genus one in the topological expansion which is proportional to $N^0$. We confirm that this term contains a dependence of the type  $\sqrt{\lambda}$, as expected from the form of the free energy of ABJM on $S^3$. In the special case when all the fugacities take the same value $\Delta_a=\pi/2$ we find a term of the form 
\be
\frac{2\pi}{3}\sqrt{2\lambda}.
\ee
The analogous term in the expansion of the free energy on $S^3$ was interpreted as a nonperturbative instanton effect in IIA string theory \cite{Drukker:2011zy}.  At the same level in the genus expansion, $N^0$, we also find a logarithmic in $\lambda$ term whose coefficient is estimated to be $-7/6$. It would be quite interesting to understand this contribution from the point of view of the supergravity one-loop effective action. 

The rest of the manuscript is organized as follows. We briefly review the topologically twisted index of ABJM theory in section \ref{Sec:Review}. Section \ref{Sec:TTI-tHooft} contains various aspects of the evaluation of the topologically twisted index in the 't Hooft limit, we consider the behavior of the eigenvalues in  great detail and evaluate the leading part of the index  using analytic techniques and present various numerical results for the sub-leading behavior. We work  in details the special case $\Delta_a=\pi/2$, because this choice has a number of simplifying properties. We also consider more generic values of $\Delta_a$. In section \ref{Sec:Gravity} we briefly discuss the gravity side. We conclude in section \ref{Sec:Conclusions}. We relegate a number of more technical issues to two appendices where we discuss, in particular, aspects of our numerical algorithm and details of the supergravity solution. 

\section{Review of the topologically twisted index of ABJM theory}\label{Sec:Review}

We follow  the works of \cite{Benini:2015noa,Closset:2016arn,Closset:2017zgf,Closset:2015rna} in general and stay particularly close to the presentation by Benini, Hristov and Zaffaroni  in  \cite{Benini:2015eyy}. The outcome of the localization process is an expression for the topologically twisted index for ABJM which we aim to numerically explore in the 't Hooft limit.

Let us briefly recall the key ingredients as a way to also explain the notation. One considers  the field theory on $S^2\times S^1$ with a background field $A^R$.  For  ABJM one includes  a set of flavor symmetries characterized by Cartan-valued magnetic background flavor symmetry
 $\frac{1}{2\pi}\int_{S^2}F^f = \vec{n}$. With these flavor symmetries one associates fugacities according to  $y=e^{i(A_t^f+i\beta \sigma^f)}$; a similar expression expression holds for the dynamical fields $x=e^{i(A_t+i\beta \sigma)}$ where the constant potential $A_t^f$ is a flat connection for the flavor symmetry and $\sigma^f$ is a real mass for the three-dimensional field theory. 

For a generic 3d ${\cal N}=2$ theory  the topologically twisted index takes the form
\be
\label{Eq:3dN2}
Z(\vec{n}, y)=\frac{1}{|W|}\sum\limits_{m\in \Gamma_{\mathfrak h}}\oint\limits_{\cal C}Z_{int}(x,y;m,\vec{n}).
\ee
The sum is over all the magnetic fluxes $m$ in the co-root lattice $\Gamma_{\mathfrak h}$ of the gauge group and the integration  over the contour ${\cal C}$. In localization, the building blocks of $Z_{int}$ are obtained from the classical action and the one-loop evaluation of determinants resulting from quantum fluctuations  around the localization locus. A chiral multiplet contributes a one-loop factor of the form 
\be
Z_{1-loop}^{chiral}=\prod\limits_{\rho \in {\cal R}}\left(\frac{x^{\rho/2}y^{\rho_f/2}}{1-x^\rho y^{\rho_f}}\right)^{\rho(m)+\rho_f(n)-q+1}, 
\ee
where ${\cal R}$ is the representation of the gauge group $G$, $\rho$ denotes the corresponding weights of the representation ${\cal R}$, $q$ denotes the  R-charge of the field and $\rho_f$ denotes the weights of the multiplet under the flavor symmetry. The vector multiplet contributes the following product of integrals 
\be
Z_{1-loop}^{gauge}=\prod\limits_{\alpha \in G}(1-x^\alpha)\left(id u\right)^r, 
\ee
where $r$ is the rank of the gauge group and $\alpha$ denotes roots of $G$.  We also use   $u=A_t+i\beta \sigma $ which lives on the complexified Cartan subalgebra  related to  $x=e^{iu}$.

The only classical contribution in the ABJM  case comes from the Chern-Simons term $Z_{class}^{CS}=x^{km}$, where $k$ is the Chern-Simons level and $m$ is a dynamical magnetic flux characterizing the localization locus and  living in the co-root lattice  $\Gamma_{\mathfrak h} $. These are the key ingredients of the construction but for  a more detailed description we refer the reader to the original treatment in \cite{Benini:2015eyy}.

In the appropriate basis for the Cartan of global symmetries one has flavor symmetries $J_{1,2,3}$ and  there is an additional R-symmetry $J_4$.  One denotes $n_{1,2,3}$  the fluxes and by $y_{1,2,3}$ the fugacities associated to $J_{1,2,3}$. To make expressions symmetric it is convenient to introduce also $n_4$ and $y_4$ such that 
\be
\sum\limits_{a=1}^4n_a=2, \qquad \prod\limits_{a=1}^4y_a=1.
\ee

The topologically twisted index for ABJM takes the form  \cite{Benini:2015eyy}:
\bea
\label{Eq:toptwistindex}
Z&=&\frac{1}{(N!)^2}\int\limits_{\cal C}\prod\limits_{i=1}^N
\frac{dx_i}{2\pi \, i x_i}\frac{d\tx_i}{2\pi \, i \tx_i}\prod\limits_{i\neq j}^N(1-\frac{x_i}{x_j})(1-\frac{\tx_i}{\tx_j}) \nonumber\\
&&\times \prod\limits_{i,j=1}^N\prod\limits_{a=1,2}
\left(\frac{\sqrt{\frac{x_i}{\tx_j}y_a}}{1-\frac{x_i}{\tx_j}y_a}\right)^{1-n_a}
\prod\limits_{b=3,4}
\left(\frac{\sqrt{\frac{\tx_j}{x_i}y_b}}{1-\frac{\tx_j}{x_i}y_b}\right)^{1-n_b} \nonumber \\
&&\times \prod\limits_{i=1}^N\frac{1}{e^{iB_i}-1} \prod\limits_{i=j}^N\frac{1}{e^{i\tilde{B}_j}-1}.
\eea
The contour of integration ${\cal C}$ follows the  Jeffrey-Kirwan prescription and simply picks up the poles in the last line of Equation (\ref{Eq:toptwistindex}) which are determined by the following ``Bethe Ansatz Equations''
\be
e^{iB_i}=x_i^k\prod\limits_{j=1}^N\frac{(1-y_3 \frac{\tilde{x}_j}{x_i})(1-y_4 \frac{\tilde{x}_j}{x_i})}{(1-y_1^{-1} \frac{\tilde{x}_j}{x_i})(1-y_2^{-1} \frac{\tilde{x}_j}{x_i})}=1,
\ee
and 
\be
e^{i\tilde{B}_j}=\tilde{x}_j^k\prod\limits_{j=1}^N\frac{(1-y_3 \frac{\tilde{x}_j}{x_i})(1-y_4 \frac{\tilde{x}_j}{x_i})}{(1-y_1^{-1} \frac{\tilde{x}_j}{x_i})(1-y_2^{-1} \frac{\tilde{x}_j}{x_i})}=1.
\ee
Once the solutions to these BAE are known, the final form of the topologically twisted index for ABJM is:
\be
\label{Eq:Index}
Z(y_a,n_a)=\prod_{a=1}^4 y_a^{-\frac{1}{2}N^2 n_a}\sum_{I\in BAE}\frac{1}{\det\mathbb{B}}
\frac{\prod_{i=1}^N x_i^N \tilde{x}_i^N\prod_{i\neq j}\left(1-\frac{x_i}{x_j}\right)\left(1-\frac{\tilde{x}_i}{\tilde{x}_j}\right)}{\prod_{i,j=1}^N\prod_{a=1,2}(\tilde{x}_j-y_ax_i)^{1-n_a}\prod_{a=3,4}(x_i-y_a\tilde{x}_j)^{1-n_a}},
\ee
where $y_a$ are the corresponding fugacities.  The summation is over all solutions $I$ of the ``Bethe Ansatz Equations" (BAE) $e^{iB_i}=e^{i\tilde{B}_j}=1$ modulo permutations. The two sets of variables $\{x_i\}$ and $\{\tilde x_j\}$ arise from the $U(N)_k\times U(N)_{-k}$ structure of ABJM theory.  Finally, the $2N\times 2N$ matrix $\mathbb{B}$ is the Jacobian relating the $\{x_i,\tilde x_j\}$ variables to the $\{e^{iB_i},e^{i\tilde B_j}\}$ variables
\begin{equation}
\mathbb B=\begin{pmatrix}x_l\fft{\partial e^{iB_j}}{\partial x_l}&\tilde x_l\fft{\partial e^{iB_j}}{\partial\tilde x_l}\\[4pt]
x_l\fft{\partial e^{i\tilde B_j}}{\partial x_l}&\tilde x_l\fft{\partial e^{i\tilde B_j}}{\partial\tilde x_l}\end{pmatrix}.
\end{equation}
It is convenient to introduce the chemical potentials $\Delta_a$ according to $y_a=e^{i\Delta_a}$. As discussed in \cite{Benini:2015eyy}, the general form of the eigenvalues for $k=1$ scales with $\sqrt{N}$ and is given by 
\be
u_i=iN^{1/2}\, t_i +v_i, \qquad \tilde{u}_i=iN^{1/2}t_i+\tilde{v}_i.
\ee

A remarkable result of \cite{Benini:2015eyy}  is the explicit treatment of Equation (\ref{Eq:Index}) to yield, as the leading term for $k=1$  in an expansion in $N$  
\be
{\rm Re}\log Z=-\frac{N^{3/2}}{3}\sqrt{2\Delta_1\Delta_2\Delta_3\Delta_4}\sum\limits_a\frac{n_a}{\Delta_a}.
\ee
A more comprehensive numerical analysis of the index and the subsequent sub-leading corrections in $N$ was presented in \cite{Liu:2017vll}. The  logarithmic in $N$ contribution to the topologically twisted index was matched with the appropriate eleven-dimensional supergravity one-loop quantum computation in \cite{Liu:2017vbl}.

\section{The topologically twisted index in the  't Hooft limit }\label{Sec:TTI-tHooft}

One of the  main goals of this manuscript is to compute ${\rm Re}\log Z$ in the 't Hooft limit, that is, for $N\to \infty$ while $\lambda=N/k$ is held fixed.  What follows bellow is essentially an analysis similar to the one performed in \cite{Benini:2015eyy} for the M-theory limit but with a few observations meant to setup large $N$ corrections to the ``saddle point'' computation.

Given our interest in extending the previous results of \cite{Benini:2015eyy} to the 't Hooft limit, we are going to carefully and explicitly track  the Chern-Simons level $ k$ in our manipulations.  Under the change of variables $x_i=e^{iu_i}, \tilde{x}_j=e^{i\tilde{u}_j},   y_a=e^{i\Delta_a}$ one finds that the Bethe Ansatz Equations become
\bea
\label{Eq:BAE}
0&=& k u_i +i\sum\limits_{j=1}^N\left[\sum\limits_{a=3,4}{\rm Li}_1(e^{i(\tilde{u}_j-u_i+\Delta_a)})-
\sum\limits_{a=1,2}{\rm Li}_1(e^{i(\tilde{u}_j-u_i-\Delta_a)})\right]-2\pi n_i, \nonumber \\
0&=& k\tilde{u}_j +i\sum\limits_{j=1}^N\left[\sum\limits_{a=3,4}{\rm Li}_1(e^{i(\tilde{u}_j-u_i+\Delta_a)})-
\sum\limits_{a=1,2}{\rm Li}_1(e^{i(\tilde{u}_j-u_i-\Delta_a)})\right]-2\pi \tilde{n}_j,
\eea
where $n_i$ and $\tilde{n}_j$ are integer numbers characterizing the ambiguity in solving $e^{iB_i}=e^{i\tilde{B}_j}=1$ and the above system of equations.  It is worth pointing out that the BAE are, in principle, exact equations determining the poles. Namely, they are not obtained in a large $N$ limit and are valid to generate the topologically twisted index for any  value of $N$ and any value of $k$. We are ultimately, of course, interested in the regime where a comparison with supergravity can be made but the exactness of this approach should have powerful implications for an eventual string theoretic understanding of the topologically twisted index as providing ultraviolet complete quantum gravity data.

It is worth noting, as remarked in  \cite{Benini:2015eyy}, that the above system of equations follows from the critical points of a ``Bethe potential''
\bea
\label{Eq:BethePotential}
\mathcal{V}&=&\sum\limits_{i=1}^N\left[\frac{k}{2}(\tilde{u}_i^2-u_i^2)-2\pi (\tilde{n}_i \tilde{u}_i-n_iu_i)\right]\nonumber \\
&&+ \sum\limits_{i,j=1}^N\left[\sum\limits_{a=3,4}{\rm Li}_2(e^{i(\tilde{u}_j-u_i+\Delta_a)})-
\sum\limits_{a=1,2}{\rm Li}_2(e^{i(\tilde{u}_j-u_i-\Delta_a)})\right].
\eea

\subsection{Numerical results: Eigenvalues and eigenvalue densities}

A natural way to encode the behavior of the eigenvalues is through its density for which we use the density of the  imaginary part $\rho(t)$ and that of the real part $\delta v(t)$. This transition  from discrete to continuous distribution is standard in problems involving large $N$ limits.  Our computation will end up being a hybrid of sorts in the sense that we use large $N$ technology with is ubiquitous introduction of eigenvalue densities but also rely on the exact numerical evaluation for finite $N$ and the corresponding numerical fitting. 

We start the analysis of the eigenvalues with the observation that a simple re-scaling corresponding to the case discussed in  \cite{Benini:2015eyy} provides some intuition for the potential generalization to the 't Hooft limit: $t \mapsto t/\sqrt{k}, \rho \mapsto \rho \sqrt{k}, \delta v \mapsto \delta v$. It is then easy to confirm numerically  that the behavior for small but arbitrary $k$ takes the form:
\bea
\label{Eq:leadingk}
u_i&=&i N^{1/2}\frac{t_i}{\sqrt{k}}+\frac{\pi}{k}-\frac{1}{2}\delta v(t_i), \nonumber \\
\tilde{u}_i&=&i N^{1/2}\frac{t_i}{\sqrt{k}}+\frac{\pi}{k}+\frac{1}{2}\delta v(t_i). 
\eea

The above expression already suggests a solution in the 't Hooft limit:
\bea
\label{Eq:leadinglambda}
u_i&=&i \sqrt{\lambda} t_i+\pi\frac{\lambda }{N}-\frac{1}{2}\delta v(t_i), \nonumber \\
\tilde{u}_i&=&i\sqrt{\lambda} t_i +\pi\frac{\lambda }{N}+\frac{1}{2}\delta v(t_i). 
\eea

The basic principle  for numerically solving large-scale equations like the BAE (\ref{Eq:BAE}) is  that the closer our starting point for the variables is to the final exact solution, the more probable and the faster it is to arrive at this solution through multidimensional root finding. Naturally, the strategy of solving the BAE is to  start from the solution in the M-theory limit, that is, for small values of $k$  and build up towards the solution in the IIA limit using the universal method of iteration. This method is based on the hypothesis that the eigenvalues change smoothly with the parameters in the equations which numerically translates into the fact that we should change the parameters of the BAE such as $N$ and $k$ (or $\lambda$)  with a small step length. Details of the numerical algorithm are in Appendix \ref{App:Algo}.

\subsubsection{Special case: $\Delta_a=\pi/2$}

\begin{figure}[h!]
\centering
\subfigure[Eigenvalue distribution]{
\includegraphics[scale=0.8]{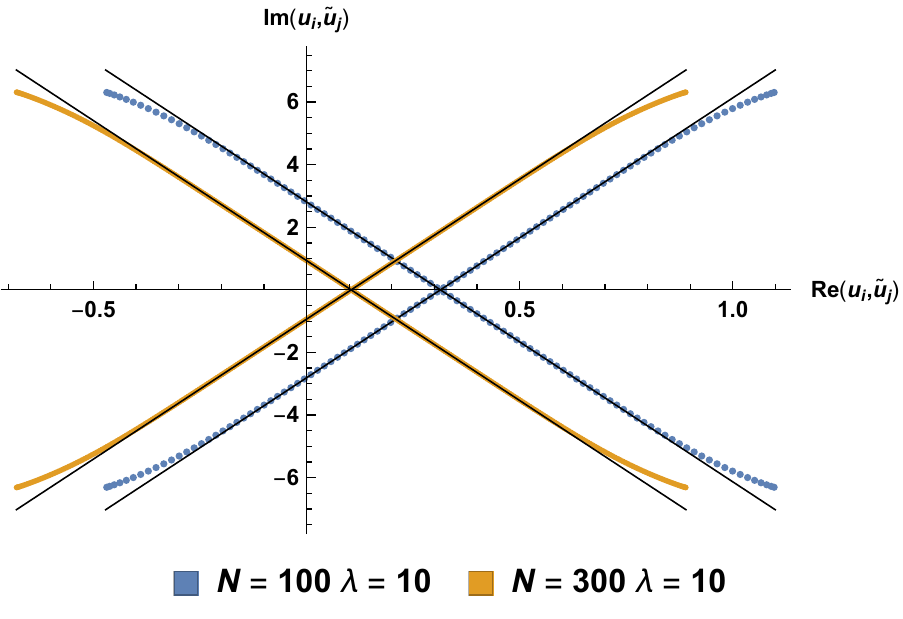}
\label{fig:Special_uubar}}
\subfigure[Eigenvalue distribution (translated to the origin)]{
\includegraphics[scale=0.8]{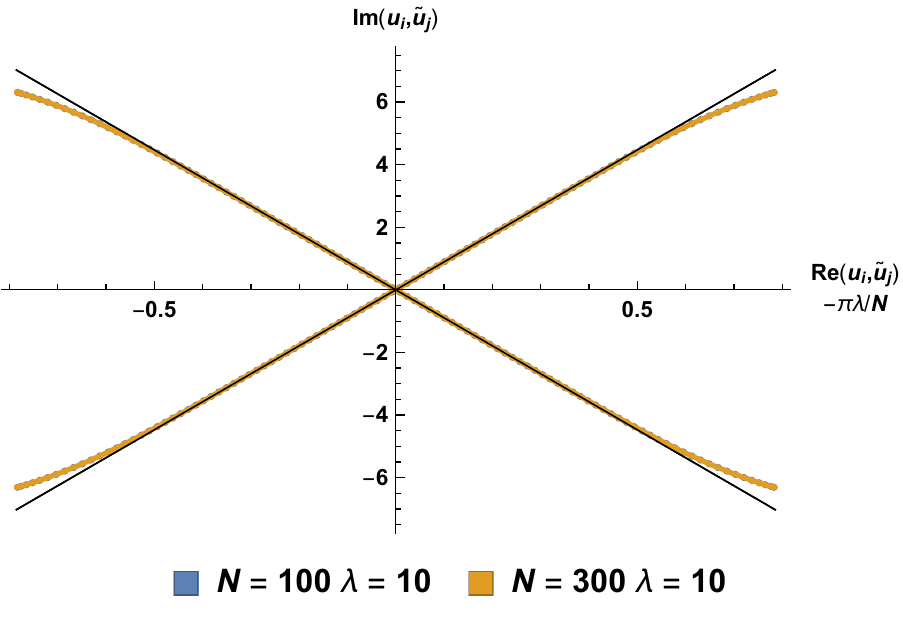}
\label{fig:Special_uubar_ori}}
\caption{Eigenvalue distribution for the special case for two different values of $N$ corresponding to $N=100$ (blue) and $300$ (orange) while keeping the same $\lambda=10$.}
\label{fig:Special_N}
\end{figure}

\begin{figure}[h!]
\centering
\subfigure[Eigenvalue density $\rho(t)$]{
\includegraphics[scale=0.8]{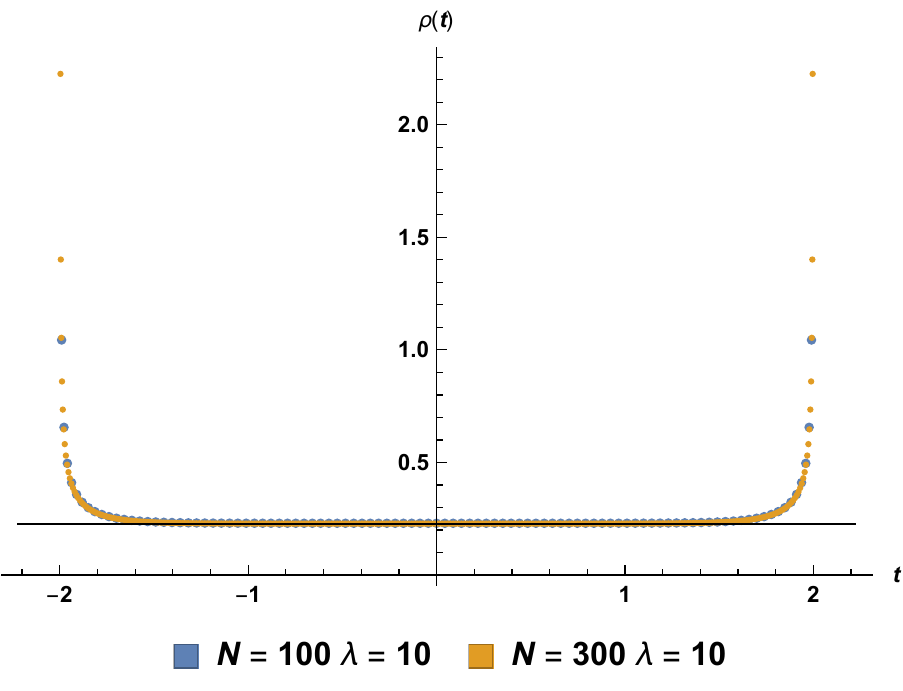}
\label{fig:Special_rho}}
\subfigure[Function $\delta v(t)$]{
\includegraphics[scale=0.8]{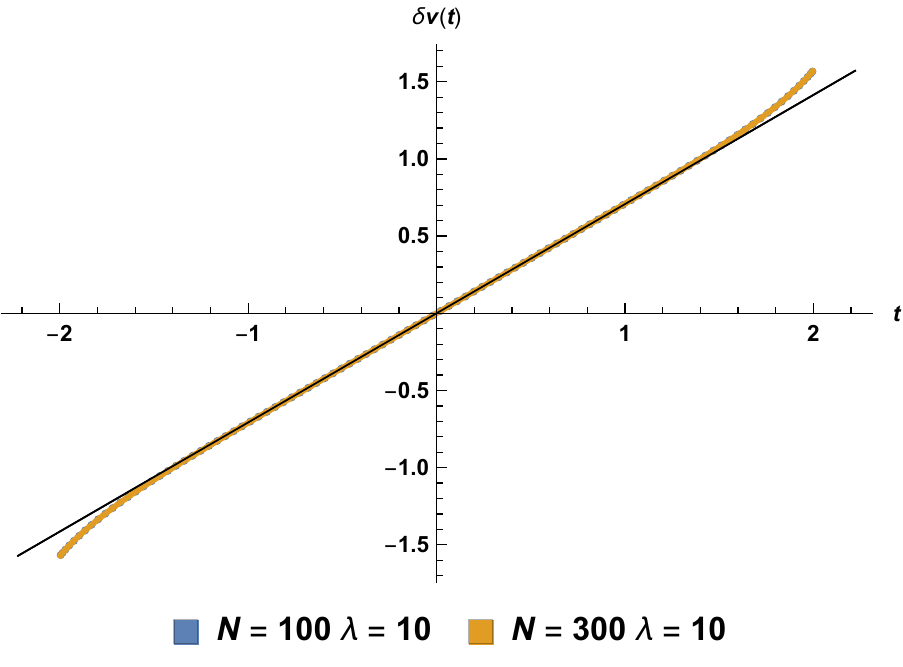}
\label{fig:Special_dv}}  
\caption{The eigenvalue density $\rho(t)$ and the function $\delta v(t)$ for the special case for $N=100$ (blue) and $300$ (orange) both for $\lambda=10$.}
\label{fig:Special_density}
\end{figure}

An example of the numerical solution for the special case (all fugacities equal) is shown in Figure \ref{fig:Special_N} and the corresponding eigenvalue density $\rho(t)$ and function $\delta v(t)$ are shown in Figure \ref{fig:Special_density}. The analytic leading order expressions for $\rho$ and $\delta v$ are the same as those in \cite{Benini:2015eyy}
\be
\rho(t)=\frac{1}{\sqrt{2}\pi},\qquad \delta v(t)=\frac{t}{\sqrt{2}},\qquad {\rm for} \ t\in\left[-\frac{\pi}{\sqrt{2}},\frac{\pi}{\sqrt{2}}\right].
\ee
These expressions correspond to the black lines in Figure \ref{fig:Special_density}. The black lines in Figure \ref{fig:Special_N} correspond to the analytical expressions in Equation (\ref{Eq:leadinglambda}).

The overlapping position of the eigenvalues as shown in Figure \ref{fig:Special_uubar_ori} as well as their continuous distributions Figure \ref{fig:Special_rho}  and \ref{fig:Special_dv} for quite different values of $N=100, 300$ but corresponding to the same $\lambda$ demonstrates numerically  that the eigenvalue densities, in this limit are independent of $N$ but depend only on $\lambda$. 

The attentive reader might have noticed that the analytic expressions, plotted as continuous black lines in the plot differ somehow from the numerical values at the endpoints of the intervals. We will discuss these deviations in the context of generic fugacities where they are more prominent.

\begin{figure}[h!]
\begin{center}
\includegraphics[scale=1.1]{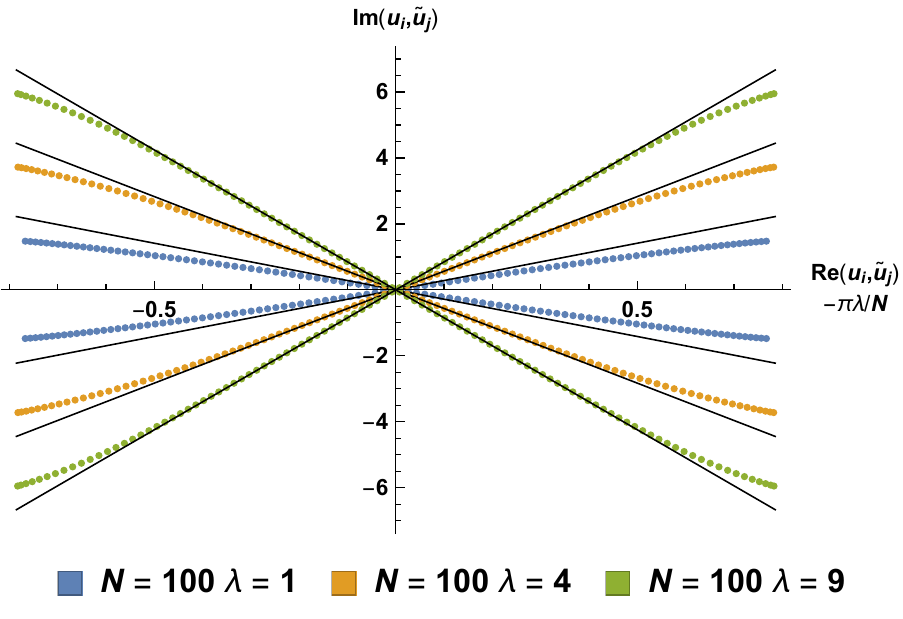}
\caption{Eigenvalue distribution (translated to the origin) for the special case for three different values of $\lambda$ corresponding to $\lambda=1$ (blue), $4$ (orange) and $9$ (green) while keeping the same $N=100$.}
\label{fig:Special_lambda_uubar_ori}
\end{center}
\end{figure}

\begin{figure}[h!]
\centering
\subfigure[Eigenvalue density $\rho(t)$]{
\includegraphics[scale=0.8]{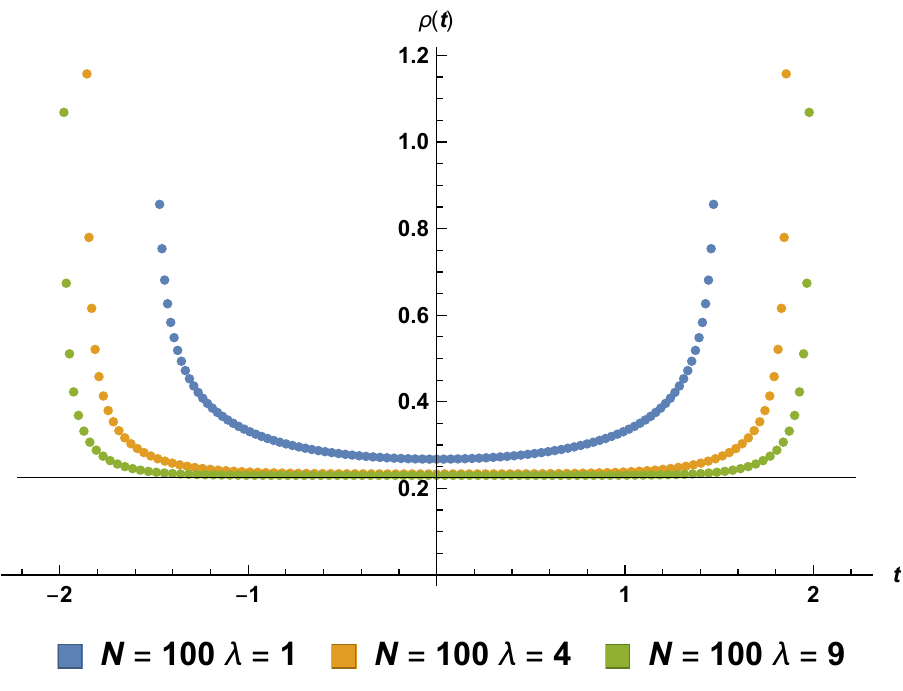}
\label{fig:Special_lambda_rho}}
\subfigure[Function $\delta v(t)$]{
\includegraphics[scale=0.8]{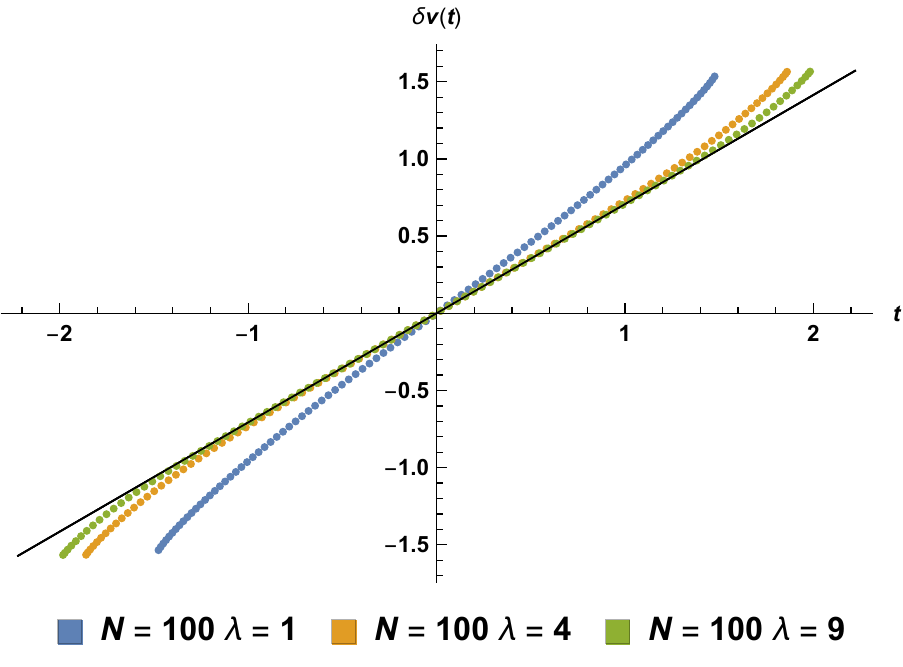}
\label{fig:Special_lambda_dv}}  
\caption{The eigenvalue density $\rho(t)$ and the function $\delta v(t)$ for the special case  for $\lambda=1$ (blue), $4$ (orange) and $9$ (green) all for $N=100$.}
\label{fig:Special_lambda_density}
\end{figure}

Having understood that the eigenvalue distribution depends only on $\lambda$ we proceed to demonstrate numerically that the scaling of the imaginary part is as $\sqrt{\lambda}$. In Figure \ref{fig:Special_lambda_uubar_ori} and Figure \ref{fig:Special_lambda_density} we plot the eigenvalue distribution, the eigenvalue density $\rho(t)$ and the function $\delta v(t)$ for various values of $\lambda=1,4,9$ for a given $N=100$. It can be  seen that the scaling is consistent with a behavior $\sqrt{\lambda}$. Note that the numerical results approach the black lines (analytical results) in  Figure
\ref{fig:Special_lambda_uubar_ori} and Figure \ref{fig:Special_lambda_density} as one increases the value of $\lambda$.  We thus expect that the agreement with gravity becomes better precisely in this regime of large $\lambda$. We will explicitly elucidate this effect when fitting the full topologically twisted index. 

Note that the solution with all equal fugacities has $\rho$ constant which points to similarities with computations in the matrix model limit of ABJM free energy on $S^3$ \cite{Herzog:2010hf}. Such similarities between  the behavior of the free energy and the topologically twisted index were noted in the M-theory limit previously in, for example, \cite{Hosseini:2016tor,Hosseini:2016ume}. We now see that there is a natural generalization of such relations in the 't Hooft limit as well. 

\subsubsection{General case: generic  $\Delta_a$}
\begin{figure}[htp!]
\centering
\subfigure[Eigenvalue distribution]{
\includegraphics[scale=0.8]{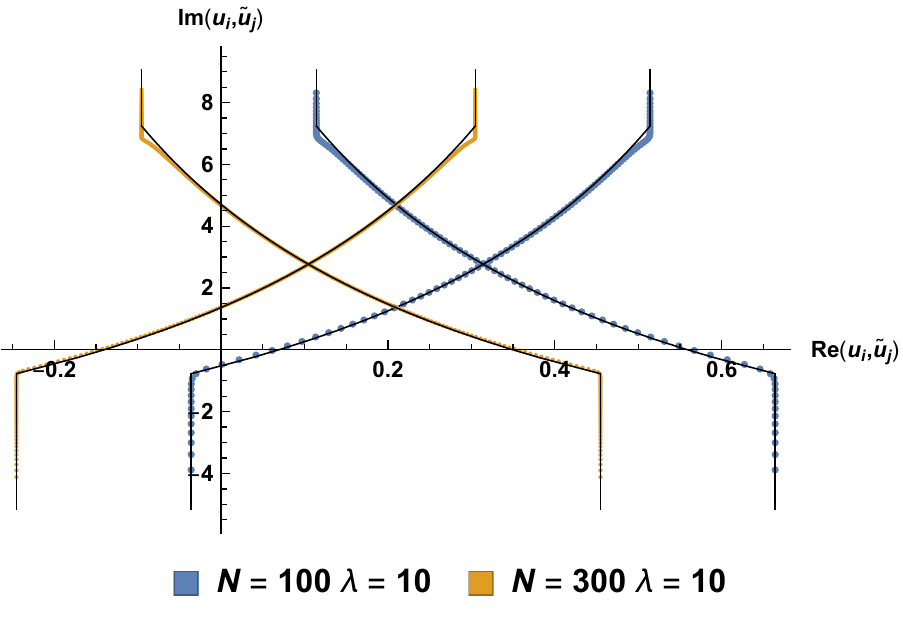}
\label{fig:457_uubar}}
\subfigure[Eigenvalue distribution (translated to the origin)]{
\includegraphics[scale=0.8]{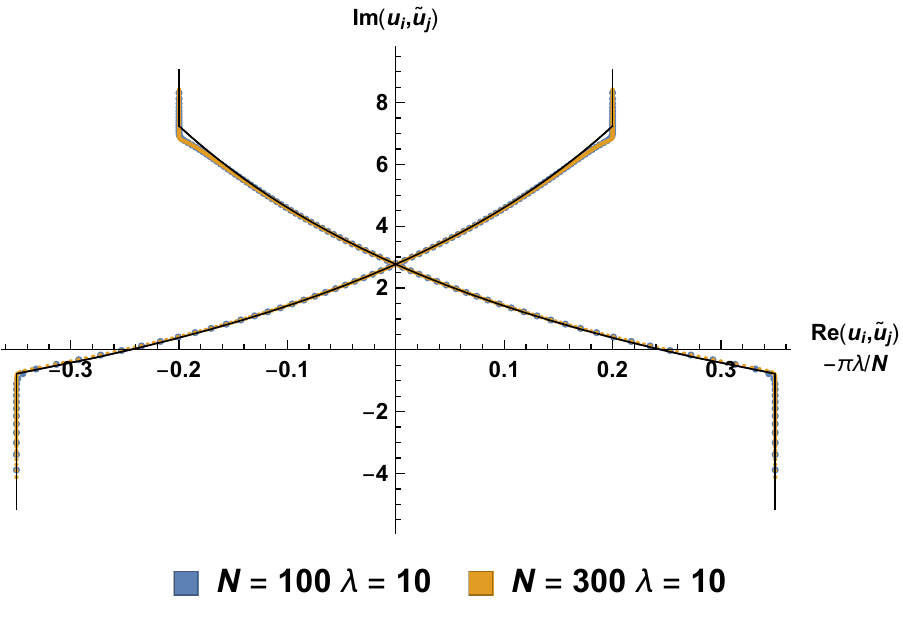}
\label{fig:457_uubar_ori}}
\caption{Eigenvalue distribution for $\Delta_a=\{0.4, 0.5, 0.7,2\pi-1.6\}$ for two different values of $N$ corresponding to $N=100$ (blue) and $300$ (orange) while keeping the same $\lambda=10$.}
\label{fig:457_N}
\end{figure}

\begin{figure}[htp]
\centering
\subfigure[Eigenvalue density $\rho(t)$]{
\includegraphics[scale=0.8]{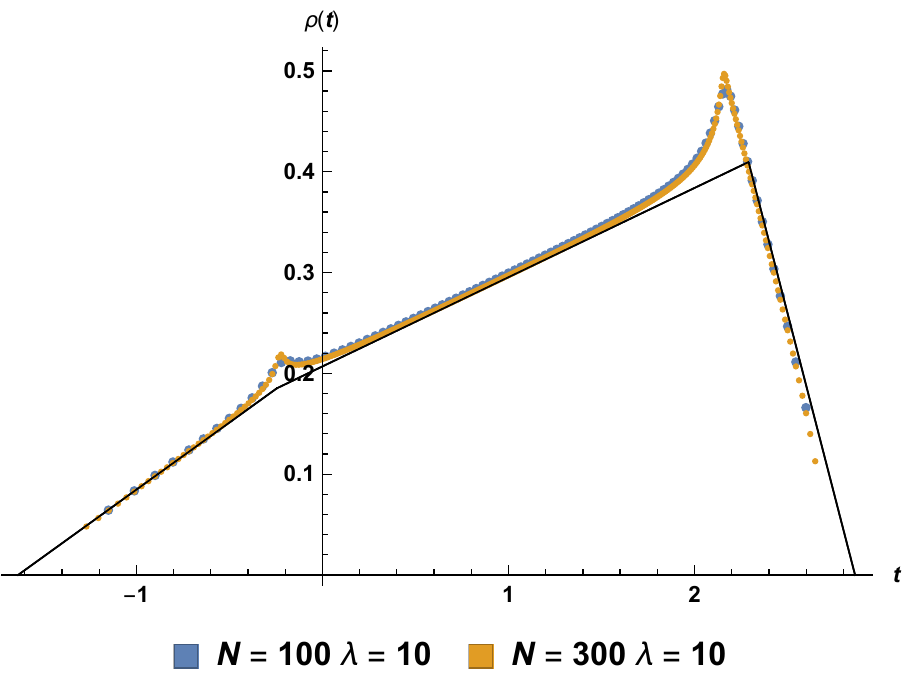}
\label{fig:457_rho}}
\subfigure[Function $\delta v(t)$  ]{
\includegraphics[scale=0.8]{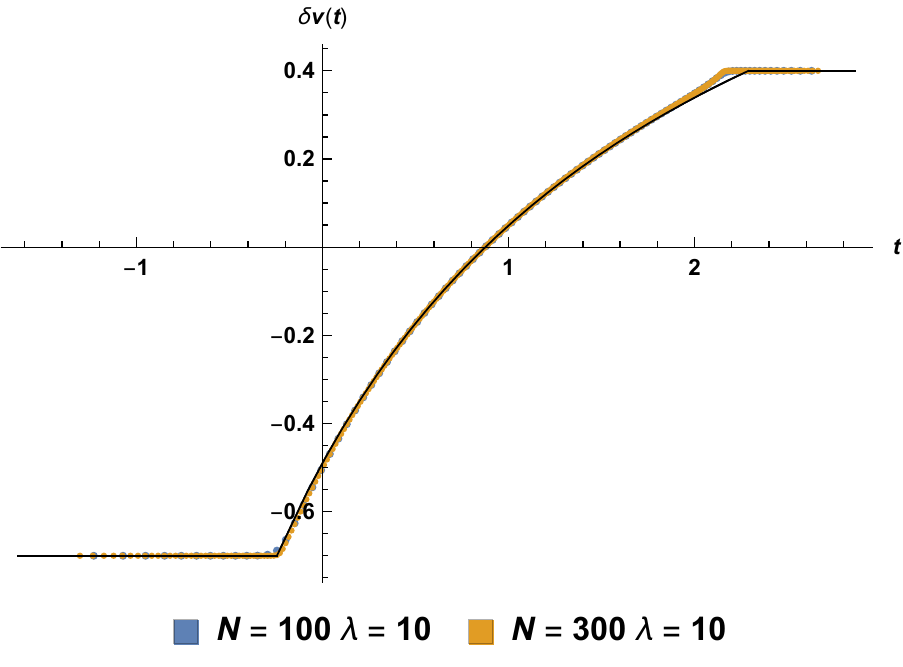}
\label{fig:457_dv}}  
\caption{The eigenvalue density $\rho(t)$ and the function $\delta v(t)$ for $\Delta_a=\{0.4, 0.5, 0.7,2\pi-1.6\}$ for $N=100$ (blue) and $300$ (orange) both for $\lambda=10$.}
\label{fig:457_density}
\end{figure}

The special case is particularly well behaved numerically. Below we present analogous results for the case of generic values of the fugacities $\Delta_a$. The conclusions are the same as those drown in  the special case. The numerical solution for $\Delta_a=\{0.4, 0.5, 0.7,2\pi-1.6\}$ is shown in Figure \ref{fig:457_N} and the corresponding eigenvalue density $\rho(t)$ and function $\delta v(t)$ are shown in Figure \ref{fig:457_density}. The overlapping plots demonstrate numerically that the structure of the eigenvalues coincides for a given value of $\lambda$ and are independent of $N$ in the large $N$ limit. 

\begin{figure}[h!]
\begin{center}
\includegraphics[scale=1.1]{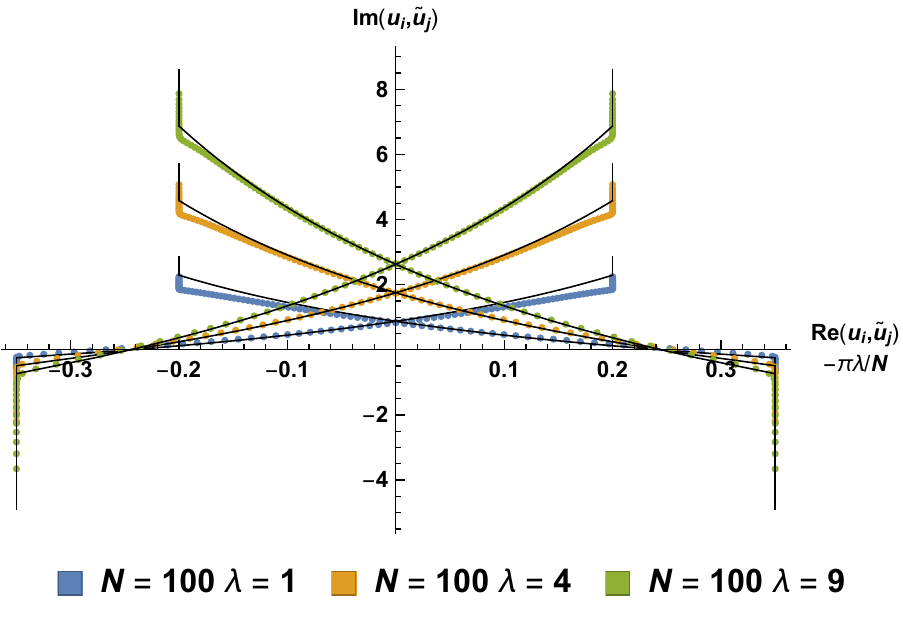}
\caption{Eigenvalue distribution (translated to the origin) for $\Delta_a=\{0.4, 0.5, 0.7,2\pi-1.6\}$ for three different values of $\lambda$ corresponding to $\lambda=1$ (blue), $4$ (orange) and $9$ (green) while keeping the same $N=100$.}
\label{fig:457_lambda_uubar_ori}
\end{center}
\end{figure}

\begin{figure}[h!]
\centering
\subfigure[Eigenvalue density $\rho(t)$]{
\includegraphics[scale=0.8]{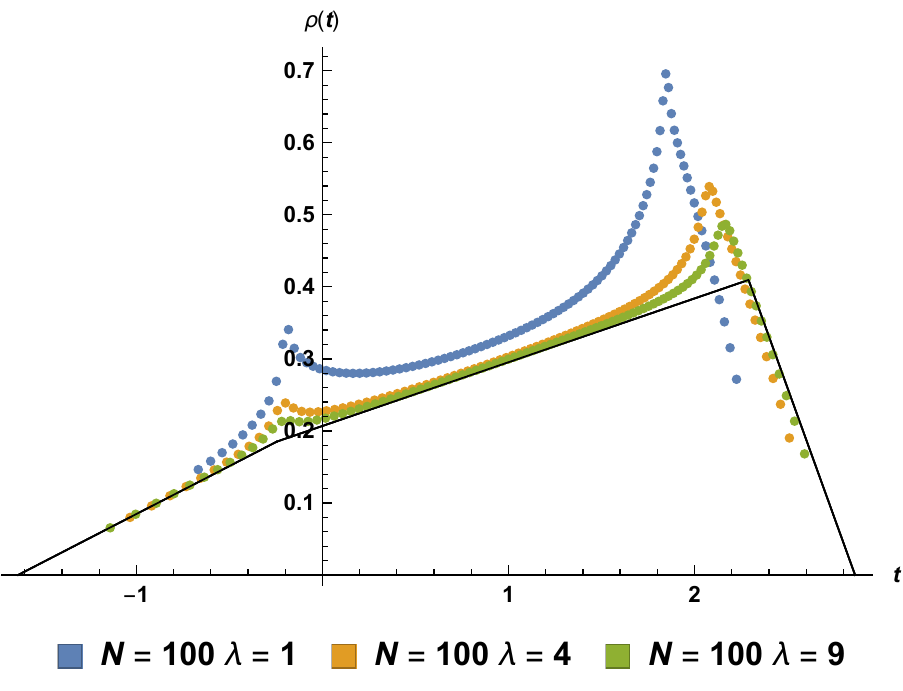}
\label{fig:457_lambda_rho}}
\subfigure[Function $\delta v(t)$]{
\includegraphics[scale=0.8]{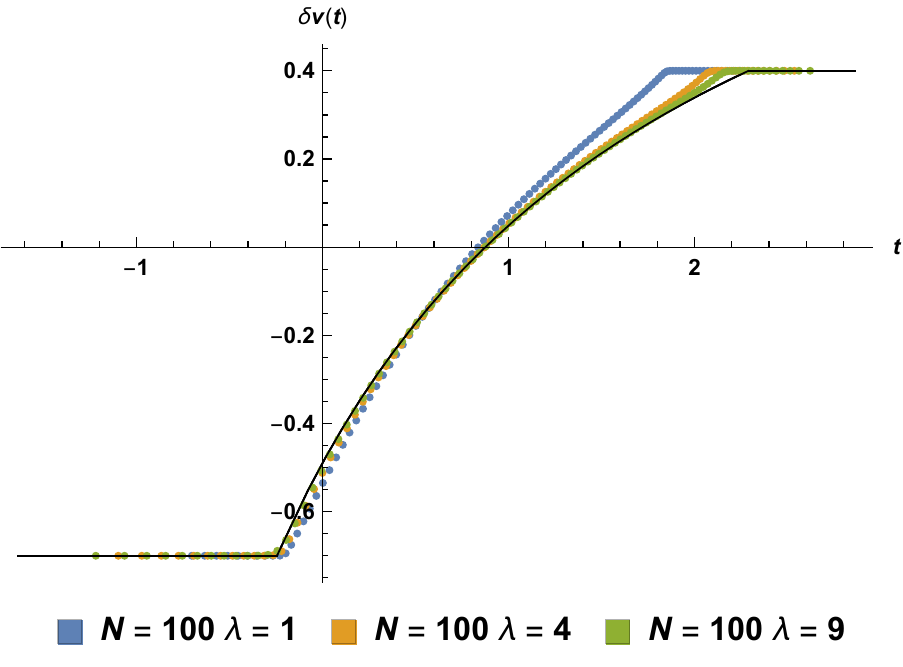}
\label{fig:457_lambda_dv}}  
\caption{The eigenvalue density $\rho(t)$ and the function $\delta v(t)$ for $\Delta_a=\{0.4, 0.5, 0.7,2\pi-1.6\}$ for $\lambda=1$ (blue), $4$ (orange)  and $9$ (green) all for $N=100$.}
\label{fig:457_lambda_density}
\end{figure}

The scaling of the imaginary part with $\sqrt{\lambda}$ and the trend that larger values of $\lambda$ lead to a better matching between the numerical result and the analytic leading expression are also on display  in Figure \ref{fig:457_lambda_uubar_ori} and Figure \ref{fig:457_lambda_density}. The effect on the tails seems to be clearly dominated by $\lambda$ and, again, becomes smaller the larger the value of $\lambda$. Figure \ref{fig:457_lambda_uubar_ori} and Figure \ref{fig:457_lambda_density} are all plotted for $N=100$ and we appreciate, even with the naked eye, the reduction of the effect of the tails as $\lambda$ increases. 

\subsection{The index }

In this section we perform a fitting of our results. The leading part of the index can be computed analytically based on the structure of the eigenvalue densities.  Having established the behavior Equation (\ref{Eq:leadinglambda}) of the eigenvalue distribution in the large $N$, large $\lambda$ limit we can proceed to analytically evaluate the topologically twisted index in the 't Hooft limit. Our starting point is the Bethe potential Equation (\ref{Eq:BethePotential}). We evaluate various parts of $\mathcal{V}$ largely following  the techniques employed in \cite{Benini:2015eyy}. Consider the first term in Equation (\ref{Eq:BethePotential}) in the large $N$ limit  it becomes
\bea
\frac{N}{2\lambda}\sum\limits_{i=1}^N(\tilde{u}_i^2-u_i^2)&=& \frac{N}{2\lambda}\sum\limits_{i=1}^N\delta v_i (2i \sqrt{\lambda} t_i +\frac{2\pi \lambda}{N}) \nonumber\\
&=& \frac{i N^2}{\sqrt{\lambda}}\int dt \rho(t) \;t\; \delta v(t) + {\cal O}(N).
\eea
Considering another term one can show that in the large $N$ limit 
\bea
\label{Eq:Partial}
\sum\limits_{i<j}{\rm Li}_2(e^{i(\tilde{u}_j-u_i-\Delta)})& \mapsto & N^2 \int dt \rho(t) \int\limits_t dt' \rho(t') {\rm Li}_2(e^{i (\tilde{u}(t')-u(t)+\Delta)}).
\eea
Following \cite{Benini:2015eyy} (see also \cite{Herzog:2010hf}) we study the behavior of this term using its expansion, namely, recall that 
\be
{\rm Li}_n (e^{iu})= \sum\limits_{m=1}^\infty\frac{e^{im u}}{m^n}.
\ee

We approximate part of the integral appearing in Equation (\ref{Eq:Partial}) using a Taylor expansion and then noticing that each subsequent term is ultimately suppressed by $1/\sqrt{\lambda}$. Namely, 
\bea
I_m (t)&=&\int\limits_t dt' \rho(t') e^{im(\tilde{u}(t')-u(t)+\Delta)}\nonumber \\
&=& \int\limits_t e^{-m\sqrt{\lambda}(t'-t)}\sum\limits_{j=0}^\infty\frac{ (t'-t)^j}{j!} \partial_x^j \big[\rho(x)e^{im\left(\frac{\delta v(x)}{2} +\frac{\delta v(t)}{2} +\Delta\right)}\big]|_{x=t} \\
&=& \frac{1}{\sqrt{\lambda}}\; \frac{\rho(t) }{m} e^{im(\delta v(t)+\Delta)}+{\cal O}(\lambda^{-1}),
\label{Eq:largelambda}
\eea
where in the second line we have introduced a Taylor expansion around $t'=t$ and in the last line we have taken the large $\lambda$ limit. 

Therefore we conclude that  in the large $N$ and further large $\lambda$  limit, the leading contribution is of the form:
\bea
\label{Eq:LargeNsum}
\sum\limits_{i<j}{\rm Li}_2(e^{i(\tilde{u}_j-u_i+\Delta)})& \mapsto & \frac{N^2}{\sqrt{\lambda}} \int dt \rho^2(t)  
{\rm Li}_3(e^{i (\delta v(t)+\Delta)}).
\eea
Similar types of manipulations in the Bethe potential lead to one of the main results of this section, the fact that our Bethe potential can be written as 
\bea
\label{Eq:BAPotential}
\mathcal{V}&=&\frac{i\, N^2}{\sqrt{\lambda}}\int dt\bigg[ t\, \rho(t) \delta v (t) + \rho^2(t)\bigg(\sum\limits_{a=3,4} g_+(\delta v(t)+\Delta_a)
-\sum\limits_{a=1,2}g_-(\delta v(t)-\Delta_a)\bigg)\bigg]\nonumber \\
&&-\mu \bigg[\int dt \rho(t) -1\bigg],
\eea
where 
\bea
g_{\pm}(u)=\frac{1}{6}u^3\mp \frac{\pi}{2}u^2+\frac{\pi^2}{3}u,
\eea
and $\mu$ is a Lagrange multiplier enforcing the normalization of the eigenvalue density. 
This is precisely the same Bethe potential discussed in \cite{Benini:2015eyy} except for the overall  factor, here $N^2/\sqrt{\lambda}$ while there $N^{3/2}$. Some preliminary analysis in \cite{Liu:2017vll} showed that, as expected, for small values of $k$, the scaling is indeed $N^{3/2}k^{1/2}$. The main conclusion following  Equation (\ref{Eq:BAPotential}) and its similarity with the Bethe potential in the M-theory  limit of the potential in \cite{Benini:2015eyy}  is that, therefore, all the extremization procedure is precisely as in \cite{Benini:2015eyy}. One can also refer to the more general analysis connecting  Bethe potentials to topologically twisted indices \cite{Hosseini:2016tor} to conclude that to leading order one has
\be\boxed{
{\rm Re}\log Z_0=-\frac{1}{3}\frac{N^2}{\sqrt{\lambda}}\sqrt{2\Delta_1\Delta_2\Delta_3\Delta_4}\sum\limits_a\frac{n_a}{\Delta_a}.}
\label{Eq:logZ0}
\ee
What we need to keep in mind is that in the 't Hooft limit, our result also uses the large $\lambda$ limit. Indeed, there are potential corrections to the leading term of the order ${\cal O}(N^2/\lambda)$. This is best seen in Equation (\ref{Eq:largelambda}) and clarifies the deviation between the numerical and analytic results for small values of lambda that was glanced already in the analysis of eigenvalues depicted in Figure \ref{fig:457_lambda_uubar_ori} and \ref{fig:457_lambda_density}. The deviation was largest for the smallest value of $\lambda$ and decreased as $\lambda$ was increased.

\subsubsection{Potential corrections to the leading order: $N^2/\sqrt{\lambda}$ vs. $N^2/\lambda$}

To better understand the potential corrections to the  leading order term let us revisit the above computation of the Bethe potential. One crucial step in taking the large $N$ limit is the evaluation in Equation (\ref{Eq:LargeNsum}). Let us repeat the derivation with more details here:
\bea
\sum\limits_{i<j}{\rm Li}_2(e^{i(\tilde{u}_j-u_i+\Delta)})&\mapsto & N^2\int dt \rho(t) \int\limits_t dt' \rho(t'){\rm Li}_2(e^{i(\tilde{u}(t')-u(t)+\Delta)}) \nonumber \\
&=&N^2\int dt \rho(t) \sum\limits_{m=1}^\infty \frac{ I_m(t)}{m^2},
\eea
where 
\bea
I_m(t)&=&\int\limits_t dt' e^{-m\sqrt{\lambda}(t'-t)}\sum\limits_{j=0}^\infty \frac{(t'-t)^j}{j!}\partial_x^j\left[ \rho(x)e^{im\left(\frac{\delta v(x)}{2} +\frac{\delta v(t)}{2} +\Delta\right)}\right]|_{x=t} = \sum\limits_{j=0}^\infty I_m^{(j)}(t), \nonumber \\
I_m^{(j=0)}&=& \frac{1}{m\sqrt{\lambda}}\rho(t)e^{im(\delta v(t)+\Delta)}, \nonumber \\
I_m^{(j=1)}&=& \frac{1}{(m\sqrt{\lambda})^{2}}\partial_x\left[ \rho(x)e^{im\left(\frac{\delta v(x)}{2} +\frac{\delta v(t)}{2} +\Delta\right)}\right]|_{x=t}\nonumber \\
&=& \frac{1}{m^2\,\lambda}\rho'(t)e^{im(\delta v(t)+\Delta)}+ \frac{i}{2m\,\lambda}\rho(t)\delta v'(t)e^{im(\delta v(t)+\Delta)},\nonumber \\
I_m^{(j\geq2)}&=& \frac{1}{(m\sqrt{\lambda})^{j+1}}\partial_x^j\left[ \rho(x)e^{im\left(\frac{\delta v(x)}{2} +\frac{\delta v(t)}{2} +\Delta\right)}\right]|_{x=t}.
\eea

Therefore:
\bea
&&\sum\limits_{i<j}{\rm Li}_2(e^{i(\tilde{u}_j-u_i+\Delta)})\mapsto N^2\int dt \rho(t) \sum\limits_{m=1}^\infty \frac{1}{m^2}\sum\limits_{j=0}^\infty I_m^{(j)}(t) \nonumber \\
&=&\frac{N^2}{\sqrt{\lambda}}\int dt \rho(t)^2 {\rm Li}_3(e^{i(\delta v(t)+\Delta)}) \nonumber \\
&&+ \frac{N^2}{\lambda}\int dt \rho(t)\left[\rho'(t){\rm Li}_4(e^{i(\delta v(t)+\Delta)})+\frac{i}{2}\rho(t)\delta v'(t){\rm Li}_3(e^{i(\delta v(t)+\Delta)})\right] \nonumber \\
&&+ \sum\limits_{j=2}^\infty \frac{N^2}{\lambda^{(j+1)/2}}\int dt \rho(t) \sum\limits_{m=1}^\infty \frac{1}{m^{j+3}}\partial_x^j\left[ \rho(x)e^{im\left(\frac{\delta v(x)}{2} +\frac{\delta v(t)}{2} +\Delta\right)}\right]|_{x=t}. 
\label{Eq:leadingcorrection}
\eea

Given that the distributions $\rho(t)$ and $\delta v(t)$ are piece-wise linear functions at the leading order, we learn from the above expansion that the sub-leading contributions are more pronounced in the neighborhoods where the functions change their slopes.  A similar behavior was noted in the M-theory limit treatment \cite{Benini:2015eyy,Liu:2017vll} and dubbed contributions from the tails; here, as indicated numerically in Figure \ref{fig:457_lambda_uubar_ori} and \ref{fig:457_lambda_density}, the controlling parameter is $1/\sqrt{\lambda}$ and makes it harder  to numerically suppress the contribution from the tails. 

\subsubsection{Free energy in the 't Hooft limit of ABJM}

A good model for the sub-leading structure of the index can be inferred from the behavior of the free energy on $S^3$. 
The 't Hooft limit of the free energy on $S^3$ for the  ABJM theory has been worked out in a series of publications, see, for example \cite{Drukker:2010nc,Drukker:2011zy,Marino:2009jd}. The main result of those publications is that the $1/N$ expansion of the free energy takes the form
\bea
F(\lambda, g_s)= \sum\limits_{g=0}^\infty g_s^{2g-2} \, F_g (\lambda), 
\eea
where 
\be
g_s=\frac{2\pi i}{k}, \quad \lambda =\frac{N}{k}.
\ee

To be completely precise $k$ is not to appear at all in the above expression and the appropriate way to read it is:
\bea
\label{Eq:TopExp}
F(\lambda, N)= \sum\limits_{g=0}^\infty \left(\frac{2\pi\, i\, \lambda}{N}\right)^{2g-2} \, F_g (\lambda).
\eea

The explicit dependence on $\lambda$ in $F_g(\lambda)$ is best stated through a variable $\kappa$ such that 
\bea
\lambda(\kappa)&=&\frac{\kappa}{8\pi}{}_3F_2\left(\frac{1}{2},\frac{1}{2},\frac{1}{2};1,\frac{3}{2};-\frac{\kappa^2}{16}\right),
\eea
where ${}_3F_2 $ is the generalized hypergeometric function. 

The quantity we are interested at the sub-leading order, $F_1(\lambda)$, is naturally written in terms of another variable $\tau$  which is related to $\kappa$ as follows:
\be
\tau =i\frac{K'(\frac{i\kappa}{4})}{K(\frac{i\kappa}{4})},
\ee
where $K$ is the  complete elliptic integral of the first kind. In this new variable $\tau$ we have 
\be
F_1(\lambda(\tau))=-\log \eta(\tau),
\ee
where $\eta(\tau)$ is the Dedekind eta function.

The expansion for the free energy $F(\lambda, N)$ is to be understood as taking very large $N$ first, and then considering its dependence on $\lambda$. In the strong coupling regime $\lambda\rightarrow\infty$ or $\kappa\rightarrow\infty$, it is convenient to use the shifted variable for relationship between $\lambda$ and $\kappa$:
\be
\hat{\lambda}=\lambda-\frac{1}{24}=\frac{1}{2\pi^2}\log^2 \kappa +{\cal O}(\kappa^{-2}),\qquad\kappa\gg1.
\label{Eq:shiftlambda}
\ee
For very large $N$ the two leading terms are $F_0(\lambda)$ and $F_1(\lambda)$ corresponding  in the topological expansion Equation (\ref{Eq:TopExp}) to genus $g=0$ and genus $g=1$, respectively. At strong coupling we get:
\bea
F_0(\lambda)&=& \frac{4\pi^3\sqrt{2}}{3}\hat{\lambda}^{3/2}+{\cal O}(e^{-2\pi\sqrt{2\hat{\lambda}}}),\nonumber \\
F_1(\lambda)&=& \frac{1}{6}\log\kappa -\frac{1}{2} \log\big[\frac{2\log \kappa}{\pi}\big]+{\cal O}(\kappa^{-2}) \nonumber \\
&=& \frac{\pi }{6}\sqrt{2\hat{\lambda}}-\frac{1}{4}\log\hat{\lambda} -\frac{3}{4}\log 2 +{\cal O}(e^{-2\pi\sqrt{2\hat{\lambda}}}).
\eea
For $g\geq0$, the leading, strong coupling behavior is given by \cite{Drukker:2011zy}:
\be
F_g(\lambda)\sim\lambda^{\frac{3}{2}-g},\qquad\lambda\rightarrow\infty, g\geq0.
\ee
Note that to obtain the contribution to the actual free energy we need to recall that this term comes multiplied by $g_s^{-2}$. If we neglect the small shift $-1/24$ for large $\lambda$, the leading term is
\bea
F_{\rm Leading} &= & g_s^{-2}F_0(\lambda)= g_s^{-2} \frac{4\pi^3\sqrt{2}}{3}\lambda^{3/2}= \left(\frac{2\pi\, i \lambda}{N}\right)^{-2} \frac{4\pi^3\sqrt{2}}{3}\lambda^{3/2} \nonumber \\
&=& -\frac{\pi \sqrt{2}}{3}\frac{N^2}{\sqrt{\lambda}}.
\label{Eq:freeEleading}
\eea
This is the quantity that was shown to perfectly match the on-shell action for supergravity on AdS$_4\times \mathbb{CP}^3$ \cite{Drukker:2010nc}. We have already shown here that the scaling agrees with the leading behavior Equation (\ref{Eq:logZ0}) of the topologically twisted index in the 't Hooft limit. 

To develop some intuition into the sub-leading behavior we might expect for the index we turn to the analogous quantity for the free energy, $F_1(\lambda)$. Note that $F_1 $ contributes to the free energy with no powers of $g_s$ as corresponds to the  genus one term. Therefore the leading and sub-leading contribution to the free energy are:
\bea
F_{\rm Leading \,\,and \,\,Subleading}&=&-\frac{\pi \sqrt{2}}{3}\frac{N^2}{\sqrt{\lambda}}\nonumber \\
&&+ N^{0}\bigg[\frac{\pi }{6}\sqrt{2\lambda}-\frac{1}{4}\log \lambda -\frac{3}{4}\log 2 \bigg] \nonumber \\
&&+\sum\limits_{g=2}^\infty \left(\frac{2\pi\, i\, \lambda}{N}\right)^{2g-2} \, F_g (\lambda),
\label{Eq:freeE}
\eea
where
\be
\sum\limits_{g=2}^\infty \left(\frac{2\pi\, i\, \lambda}{N}\right)^{2g-2} \, F_g (\lambda)\sim
\sum\limits_{g=2}^\infty N^{2-2g}\lambda^{g-1/2}.
\label{Eq:freeEsub}
\ee

This approximation is the crucial one and we will show how it informs our fitting results. We will fit our numerical results following the structure of the free energy. The genus expansion predicts terms of the form $N^{2-2g}, g=0,1,2,\cdots$.   The  behavior of the  $F_1$ term in the appropriate limit it leads to  a logarithmic dependence of the type 
\be
 -\frac{1}{4} \log \lambda.
\ee
We, therefore, expect an  analogous term for the topologically twisted index. 

\subsubsection{Numerical results: The leading term of the index}

Having obtained the leading order expression analytically and the guidance from the free energy on $S^3$, we now proceed to consider sub-leading contributions to the topologically twisted index of ABJM. We expand the index beyond the leading order in $N$ and we expect the sub-leading behavior of the index to have the form 
\be
{\rm Re}\log Z=f_1(\lambda,\Delta_a,n_a)N^2+f_2(\lambda,\Delta_a,n_a)\log{N}+f_3(\lambda,\Delta_a,n_a)+\mathcal O(N^{-2}),
\label{Eq:FitlogZ}
\ee
where the functions $f_1$, $f_2$ and $f_3$ are linear in the magnetic fluxes $n_a$. Our goal is to numerically clarify the structure of the functions $f_1$, $f_2$ and $f_3$.  The function $f_1$ clearly defines the leading term, the function $f_2$ deserves some explanation in the current context which we now provide.  It is well known \cite{Ooguri:2002gx} that there are terms that can be present in the exact  expression for certain partition functions but are not captured in the 't Hooft limit. This is typical in the context of Chern-Simons theories as explained in detail in \cite{Ooguri:2002gx}; the precise reason being the residual global  gauge symmetries  corresponding to constant $U(N)$  gauge transformations. As a result, the non-perturbative partition function contains a volume factor which ultimately is responsible for $\log N$ terms in the exact evaluation which are not captured by the 't Hooft expansion. Our numerical computation, being exact, does indeed  contain such $\log N$ term and we take it into consideration when fitting by introducing the $f_2$ function\footnote{We thank Kazumi Okuyama for important clarifications on this point.}. 

For a given set of chemical potentials $\Delta_a$ and a fixed $\lambda$, we compute the index Equation (\ref{Eq:Index}) and its real part ${\rm Re}\log Z$ for a range of $N$. Since the solutions are ``$k$-fold degenerate'' for $k>1$, we should sum over all the orbits and multiply the index by $k=(N/\lambda)$\cite{Benini:2015eyy}. We then decompose ${\rm Re}\log Z$ into a sum of four independent terms
\be
{\rm Re}\log Z=A+B_1n_1+B_2n_2+B_3n_3,
\ee
where we have used the condition $\sum_a n_a=2$. Then we perform a linear least-squares fit for $A$ and $B_a$ to the function
\be
f(N)=f_1N^{2}+f_2\log{N}+f_3+\sum\limits_{g=2}^{g_c} f_{g+2}N^{2-2g},
\label{Eq:FitlogZN}
\ee
where $g_c$ is the cutoff value of the genus $g$. Since the largest value of $N$ is finite (about 300), it is important to consider the inverse powers of $N$. Guided by numerical stability which has been checked in various regimes, the more sub-leading terms added in the fitting, the more accurate the fitting results will be. The maximum number of the fitting terms equals the number of $N$, but the real number of the fitting terms is always less than the maximum to avoid the over-fitting problem.

The results of the numerical fit for ${\rm Re}\log Z$ with $N$ are presented in Table \ref{tbl:FitlogZDat}. The numerical results indicate  that the coefficient $f_2$ of the $\log N$ term is precisely $2/3$.

The comparison between the numerical leading term $f_1N^{2}$ and the analytical leading term ${\rm Re}\log Z_0$ for the special case is shown in Table \ref{tbl:RelogZ0}. Thus, showing that the numerical and the analytical values approach each other with increasing  $\lambda$ to a precision of the order of half a percent.

\begin{table}[h!]
\centering
\begin{tabular}{l|l|l|l||l|l|l}
$\lambda$&$\Delta_1$&$\Delta_2$&$\Delta_3$&$f_1$&$f_2$&$f_3$\\
\hline
\multirow{9}{*}{$1$}&$\pi/2$&$\pi/2$&$\pi/2$&$-1.43536$&$0.66667$&$0.82149$ \\
\cline{2-7}
&$0.3$&$0.4$&$0.5$&\tabincell{l}{$-0.13041$\\$-0.75745n_1$\\$-0.55716n_2$\\$-0.44271n_3$}&\tabincell{l}{$0.66667$\\$+2.11753\times10^{-17}n_1$\\$-8.54078\times10^{-20}n_2$\\$-8.55415\times10^{-20}n_3$}&\tabincell{l}{$4.45202$\\$-0.11447n_1$\\$-0.44612n_2$\\$-0.35008n_3$} \\
\cline{2-7}
&$0.4$&$0.5$&$0.7$&\tabincell{l}{$-0.19246$\\$-0.80669n_1$\\$-0.63097n_2$\\$-0.43713n_3$} & \tabincell{l}{$0.66667$\\$+2.30654\times10^{-20}n_1$\\$+3.85179\times10^{-23}n_2$\\$-2.71336\times10^{-23}n_3$} & \tabincell{l}{$3.43678$\\$-0.08378n_1$\\$-0.31636n_2$\\$-0.25899n_3$}  \\
\hline
\multirow{9}{*}{$5$}&$\pi/2$&$\pi/2$&$\pi/2$&$-0.65587$&$0.66667$&$2.62506$\\
\cline{2-7}
&$0.3$&$0.4$&$0.5$&\tabincell{l}{$-0.04701$\\$-0.36002n_1$\\$-0.26436n_2$\\$-0.20773n_3$}&\tabincell{l}{$0.66667$\\$+1.59689\times10^{-6}n_1$\\$+3.19791\times10^{-8}n_2$\\$-1.80238\times10^{-8}n_3$}&\tabincell{l}{$12.48909$\\$-0.52544n_1$\\$-1.23230n_2$\\$-1.18290n_3$} \\
\cline{2-7}
&$0.4$&$0.5$&$0.7$&\tabincell{l}{$-0.07381$\\$-0.38478n_1$\\$-0.30060n_2$\\$-0.20521n_3$} & \tabincell{l}{$0.66667$\\$+5.93294\times10^{-8}n_1$\\$+2.56635\times10^{-9}n_2$\\$-3.06747\times10^{-10}n_3$} & \tabincell{l}{$9.85037$\\$-0.41298n_1$\\$-0.90099n_2$\\$-0.90971n_3$}  \\
\hline
\multirow{9}{*}{$10$}&$\pi/2$&$\pi/2$&$\pi/2$&$-0.46585$&$0.66667$&$4.56578$ \\
\cline{2-7}
&$0.3$&$0.4$&$0.5$&\tabincell{l}{$-0.03264$\\$-0.25646n_1$\\$-0.18831n_2$\\$-0.14765n_3$}&\tabincell{l}{$0.66699$\\$+0.00037n_1$\\$+0.00003n_2$\\$-5.06570\times10^{-6}n_3$}&\tabincell{l}{$19.08969$\\$-0.82192n_1$\\$-1.81001n_2$\\$-1.80619n_3$} \\
\cline{2-7}
&$0.4$&$0.5$&$0.7$&\tabincell{l}{$-0.05168$\\$-0.27410n_1$\\$-0.21415n_2$\\$-0.14585n_3$} & \tabincell{l}{$0.66671$\\$+0.00004n_1$\\$+4.99553\times10^{-6}n_2$\\$-5.58231\times10^{-7}n_3$} & \tabincell{l}{$15.22590$\\$-0.64388n_1$\\$-1.32871n_2$\\$-1.39319n_3$}  \\
\end{tabular}
\caption{Numerical fit for ${\rm Re}\log Z=f_1N^{2}+f_2\log{N}+f_3+\cdots$.  The values of $N$ used in the fit range from 100 to 300 in steps of 10 for the special case and from 50 to 300 in steps of 10 for the general cases. We made use of the fact that the index is independent of the magnetic fluxes when performing the fit for the special case ($\Delta_a=\{\pi/2,\pi/2,\pi/2,\pi/2\}$).}
\label{tbl:FitlogZDat}
\end{table}

\begin{table}[h]
\centering
\begin{tabular}{c||l|l||l}
$\lambda$ & $f_1$ & ${\rm Re}\log Z_0/N^2$ & Error\\
\hline
$1$ & $-1.43536$ & $-1.48096$ & $3.177\%$ \\
\hline
$5$ & $-0.65587$ & $-0.66231$ & $0.982\%$ \\
\hline
$10$ & $-0.46585$ & $-0.46832$ &$0.530\%$ \\
\end{tabular}
\caption{The comparision between the numerical and the analytical values of the leading term for the special case.}
\label{tbl:RelogZ0}
\end{table}

Now we focus on the relationship between $f_1$ and $\lambda$. From the corrections of the leading term in Equation (\ref{Eq:leadingcorrection}), we expect the behavior of the leading term $f_1$ to have the form
\be
f_1(\lambda,\Delta_a,n_a)=g_1(\Delta_a,n_a)\frac{1}{\sqrt{\lambda}}+g_2(\Delta_a,n_a)\frac{1}{\lambda}+\mathcal O(\lambda^{-\frac{3}{2}}).
\ee
Using a similar decomposition
\be
f_1(\lambda,\Delta_a,n_a)=C+D_1n_1+D_2n_2+D_3n_3,
\ee
we perform a linear least-squares fit of $C$ and $D_a$ to the function
\be
g(\lambda)=g_1\frac{1}{\sqrt{\lambda}}+g_2\frac{1}{\lambda}+g_3\frac{1}{\lambda^{3/2}}+g_4\lambda^{-2}+\sum\limits_{j=4}^{j_c} g_{j+1}\lambda^{-(j+1)/2},
\label{Eq:FitlogZleading}
\ee
where $j_c$ is the cutoff value of $j$, which is similar to $g_c$ in Equation (\ref{Eq:FitlogZN}). It is implied in Equation (\ref{Eq:freeEsub}) that the larger value of $\lambda$, the larger error of the fitting results, because we have to make a cutoff in the fitting and retain finite sub-leading terms in $N$. Thus $\lambda$ should not be too large, especially for the general cases it is required that $\lambda \ll N$.

\begin{table}[h!]
\centering
\begin{tabular}{l|l|l||l|l|l|l}
$\Delta_1$&$\Delta_2$&$\Delta_3$&$g_1$&$g_2$&$g_3$&$g_4$\\
\hline
$\pi/2$&$\pi/2$&$\pi/2$&$-1.48096$&$6.82484\times10^{-8}$&$0.09256$&$0.04567$ \\
\hline
$0.3$&$0.4$&$0.5$&\tabincell{l}{$-0.10234$\\$-0.81680n_1$\\$-0.59953n_2$\\$-0.47071n_3$} & \tabincell{l}{$-0.00152$\\$+0.00352n_1$\\$-0.00230n_2$\\$+0.01714n_3$} & \tabincell{l}{$0.02172$\\$+0.01924n_1$\\$+0.06301n_2$\\$-0.10581n_3$}&\tabincell{l}{$-0.14090$\\$+0.19326n_1$\\$-0.15096n_2$\\$+0.65523n_3$}  \\
\hline
$0.4$&$0.5$&$0.7$&\tabincell{l}{$-0.16278$\\$-0.87298n_1$\\$-0.68142n_2$\\$-0.46418n_3$} & \tabincell{l}{$-0.00283$\\$+0.00245n_1$\\$-0.00644n_2$\\$+0.00423n_3$} & \tabincell{l}{$0.03041$\\$+0.04017n_1$\\$+0.09167n_2$\\$+0.00489n_3$}&\tabincell{l}{$-0.13613$\\$+0.06357n_1$\\$-0.21940n_2$\\$+0.05864n_3$} \\
\end{tabular}
\caption{Numerical fit for $f_1=g_1/\sqrt{\lambda}+g_2/\lambda+g_3/\lambda^{3/2}+g_4\lambda^{-2}+\cdots$. In the fit, for the special case, $\lambda$ ranges from 20 to 50 in steps of 5 and $N$ ranges from 100 to 400 in steps of 10; for the general cases, $\lambda$ ranges from 5 to 15 in steps of 1 and $N$ ranges from 50 to 250 in steps of 5.}
\label{tbl:FitlogZleadingDat}
\end{table}

The results of the numerical fit for $f_1$ with $\lambda$ are presented in Table \ref{tbl:FitlogZleadingDat}. One important numerical result is that, for the special case, the leading coefficient $g_1$ of the $N^2/\sqrt{\lambda}$ term matches the analytical value precisely in Equation (\ref{Eq:logZ0}) and (\ref{Eq:freeEleading})
\be
g_1=-1.48096=-\frac{\pi\sqrt{2}}{3}.
\ee 
For the cases of generic fugacities the leading coefficient also matches the analytical expression in Equation (\ref{Eq:logZ0}).

It is worth pointing out that the numerical value in Table \ref{tbl:FitlogZleadingDat} of the coefficient $g_2$ of the $N^2/\lambda$ term in Equation (\ref{Eq:FitlogZleading}) is vanishingly small. This is despite our original analytical estimation in Equation (\ref{Eq:leadingcorrection}). What the numerical approach is showing is that there is a sharp cancellation among the various terms contributing at $N^2/\lambda$ order. Such cancellations are possible due to the symmetries of the ABJM model, at a more technical level they are consequences of identities among various polylogarithms.  They have not been explored in the 't Hooft limit and we hope to return to an analytic proof in a separate publication. In fact, the absence of the $N^2/\lambda$ term exactly corresponds to the absence of the $\mathcal O(N)$ term in the M-theory limit in \cite{Liu:2017vll}, as we will argue later in this section. 

Finally, the numerical analysis remarkabfly  captures the shift $-1/24$ in $\lambda$ in Equation (\ref{Eq:shiftlambda}). Indeed, through Taylor expansion, the expression for the leading free energy of ABJM on $S^3$ becomes \cite{Drukker:2011zy}:
\bea
F_{{\rm Leading}}&=&g_s^{-2}\frac{4\pi^3\sqrt{2}}{3}\hat{\lambda}^{3/2}  \nonumber \\
&=& -\frac{\pi \sqrt{2}}{3}\frac{N^2}{\sqrt{\lambda}}+\frac{\pi}{24\sqrt{2}}\frac{N^2}{\lambda^{3/2}}.
\eea
This shifted result  agrees perfectly with the numerical analysis. Note that  for the special case in Table \ref{tbl:FitlogZleadingDat}, the coefficient $g_3$ of the $N^2/\lambda^{3/2}$ term is precisely
\be
g_3=0.09256=\frac{\pi}{24\sqrt{2}}.
\ee 
\subsubsection{Numerical results: The first sub-leading term of the index}
Next we explore the dependence of the first sub-leading term of the index, namely the function  $f_3$ of the $N^0$ term, on $\lambda$. From the topological expansion of the free energy on $S^3$ to genus $g=0$ and genus $g=1$ in Equation (\ref{Eq:freeE}), we expect the behavior of the function $f_3$ to adhere to the following pattern
\be
f_3(\lambda,\Delta_a,n_a)=h_1(\Delta_a,n_a)\:\sqrt{\lambda}+h_2(\Delta_a,n_a)\,\log{\lambda}+h_3(\Delta_a,n_a)+\mathcal O(e^{-\sqrt{\lambda}}).
\ee
Through the same decomposition we consider 
\be
f_3(\lambda,\Delta_a,n_a)=E+F_1n_1+F_2n_2+F_3n_3,
\ee
and perform a linear least-squares fit of $E$ and $F_a$ to the function
\be
h(\lambda)=h_1\:\sqrt{\lambda}+h_2\,\log{\lambda}\,+h_3.
\ee

The results of the numerical fit for $f_3$ with $\lambda$ are presented in Table \ref{tbl:FitlogZconstantDat}. The numerical results show that the coefficient $h_2$ of the $\log{\lambda}$ term is $-7/6$.

\begin{table}[h!]
\centering
\begin{tabular}{l|l|l||l|l|l}
$\Delta_1$&$\Delta_2$&$\Delta_3$&$h_1$&$h_2$&$h_3$\\
\hline
$\pi/2$&$\pi/2$&$\pi/2$&$2.96129$&$-1.16397$&$-2.10347$ \\
\hline
$0.3$&$0.4$&$0.5$&\tabincell{l}{$7.96583$\\$-0.30541n_1$\\$-0.61946n_2$\\$-0.66859n_3$} & \tabincell{l}{$-1.14322$\\$-0.01480n_1$\\$-0.00297n_2$\\$-0.00320n_3$} & \tabincell{l}{$-3.41833$\\$+0.17809n_1$\\$+0.15182n_2$\\$+0.31098n_3$}\\
\hline
$0.4$&$0.5$&$0.7$&\tabincell{l}{$6.65321$\\$-0.23751n_1$\\$-0.45679n_2$\\$-0.52009n_3$} & \tabincell{l}{$-1.15405$\\$-0.01422n_1$\\$-0.00479n_2$\\$-0.00048n_3$} & \tabincell{l}{$-3.11676$\\$+0.13854n_1$\\$+0.12381n_2$\\$+0.24913n_3$}\\
\end{tabular}
\caption{Numerical fit for $f_3=h_1\sqrt{\lambda}+h_2\log{\lambda}+h_3$. In the fit, for the special case, $\lambda$ ranges from 20 to 50 in steps of 5 and $N$ ranges from 100 to 400 in steps of 10; for the general cases, $\lambda$ ranges from 5 to 15 in steps of 1 and $N$ ranges from 50 to 250 in steps of 5. To be more accurate, we replace $\lambda$ with $\hat{\lambda}=\lambda-1/24$ in the fit.}
\label{tbl:FitlogZconstantDat}
\end{table}

Similar to the comparison performed in  \cite{Liu:2017vll}, for the special case we find the approximate expression
\be
h_1=2.96129\approx\frac{2\pi}{3}\sqrt{2}.
\ee
A similar term, proportional to $\sqrt{\lambda}$, appears at genus one order in the topological expansion of the free energy on $S^3$ of the ABJM theory. It was argued  that in the field theory dual it originates as instanton effects due to tunneling of eigenvalues in the respective matrix model. On the gravity side this very term was attributed to D2 brane instantons wrapping a warped $\mathbb{RP}^3$ inside $\mathbb{CP}^3$ \cite{Drukker:2011zy} (see also  \cite{Okuyama:2016deu} for a detailed treatment of instanton corrections). It is plausible that a similar origin is at play in our context of the topologically twisted index. On the gravity side, as we will see in the next section, the structure of $\mathbb{CP}^3$ is deformed due to the magnetic charges, which makes the D2 brane argument more subtle. Another promising venue to explore is along the lines of higher curvature corrections which, as shown in \cite{Liu:2010gz}, might give rise to $\sqrt{\lambda}$ contributions for theories on AdS$_4$.

Thus, the topologically twisted index in the special case when all the fugacities take the same value, $\Delta_a=\pi/2$, is approximately of the form
\be
{\rm Re}\log Z(\Delta_a=\pi/2)=-\frac{\pi\sqrt{2}}{3}\frac{N^2}{\sqrt{\lambda}} +\frac{2}{3}\log{N}+\frac{2\pi}{3}\sqrt{2\lambda}-\frac{7}{6}\log\lambda+\mathcal O(1)+\mathcal O(N^{-2}).
\label{Eq:tti}
\ee
One of our main results in this manuscript is the prediction, from the field theory side, for the term $-(7/6)\log \lambda$; this seems to be persistent and robust for generic values of the fugacities as  indicated numerically in Table  \ref{tbl:FitlogZconstantDat}. This term, as christened by Ashoke Sen in a series of papers \cite{Sen:2011ba,Sen:2012cj}, constitutes an infrared window into ultraviolet physics. Here we have obtained it by exploring the field theory side which via the AdS/CFT correspondence provides an ultraviolet complete description of the dual gravity. The very same term should be reproduced by a  one-loop computation on the gravity side using only the massless sector.  In the context of asymptotically flat black holes in string theory such logarithmic corrections have been reproduced in a variety of situations \cite{Sen:2014aja}. In asymptotically AdS spacetimes positive results have been reported in the M-theory limit \cite{Bhattacharyya:2012ye,Liu:2017vbl,Gang:2019uay}; it would be quite interesting to reproduce such logarithmic correction in the 't Hooft limit from the dual type IIA supergravity.

Let us conclude this section with an extra sanity check. In the M-theory limit it was found a logarithmic term of the form $-(1/2)\log{N}$ \cite{Liu:2017vll}. One might naturally ask how that logarithmic term fits with the terms we find in this manuscript. It turns our that adding a term of the form $(7/6)\log k$, yields the two coefficients that we have established in the 't Hooft limit: $(2/3)\log N$ and $-(7/6)\log \lambda$. In fact all of the numerical results in the 't Hooft limit shown above correspond to the results in the M-theory limit through change of variables $k\rightarrow N/\lambda$. The complete correspondence for the special case is
\bea
{\rm Re}\log Z(\Delta_a=\pi/2)&=&-\frac{\pi\sqrt{2k}}{3}N^{3/2}+\frac{\pi}{\sqrt{2k}}\left(\frac{k^2}{24}+\frac{1}{3}+1 \right)N^{1/2}\nonumber \\
&&-\frac{1}{2}\log{N}+\frac{7}{6}\log{k}+\cdots \nonumber \\
&\Longrightarrow&-\frac{\pi\sqrt{2}}{3}\frac{N^2}{\sqrt{\lambda}}+\frac{\pi}{24\sqrt{2}}\frac{N^2}{\lambda^{3/2}}+\frac{2}{3}\log{N}\nonumber \\
&&+\frac{2\pi}{3}\sqrt{2\lambda}-\frac{7}{6}\log\lambda+\cdots,
\label{Eq:correspondence}
\eea
accommodating also the absence of $N^2/\lambda$ terms discussed previously from the numerical point of view. 

\section{Dual magnetically charged asymptotically AdS$_4$ black holes}\label{Sec:Gravity}
 
 The  low energy gravity dual to the 't Hooft limit of  ABJM theory is best described by IIA supergravity  on AdS$_4\times \mathbb{CP}^3$ \cite{Aharony:2008ug}
 \bea
 \label{Eq:GravityBack}
 ds^2_{string}&=& \frac{R^3}{k}\left(\frac{1}{4}ds^2(AdS_4)+ds^2(\mathbb{CP}^3)\right), \nonumber \\
 F_4&=&\frac{3}{8}R^3\hat{\epsilon}_4, \qquad F_2=k J, \nonumber \\
 e^{2\phi}&=& \frac{R^3}{k^3},
\eea
where $\hat{\epsilon}_4$ is the volume form on AdS$_4$,  $J$ is the Kaehler form on $\mathbb{CP}^3$.  This solution is a reduction from 11d supergravity of the Freund-Rubin solution of the form AdS$_4\times S^7/\mathbb{Z}_k$ when $S^7$ is viewed as a $U(1)$ bundle over $\mathbb{CP}^3$.  The IIA viewpoint is the appropriate one for large values of $k$ for which the 11d circle would have been very small leading to a breakdown of the 11d supergravity approximation.  The radius of curvature in string units is \cite{Aharony:2008ug}
\be
R_{str}^2=\frac{R^3}{k}=2^{5/2}\pi \sqrt{\lambda}.
\ee
This expression fits nicely with the field theory expectation that contains in the genus-one term of the topological expansion, that is, in the term of the form $N^0$, terms of  the form $\sqrt{\lambda}$ and also $\log \lambda$.  Thus, we expect that, indeed, the logarithmic term might be reproduced by a one-loop supergravity computation of the effective action. Such one-loop corrections are naturally proportional to the logarithm of the overall size.


The above background in Equation (\ref{Eq:GravityBack}) corresponds to the vacuum of ABJM theory. To make direct contact with the topologically twisted index we are discussing, one needs to consider more general magnetically charged configurations. Let us start by considering the magnetically charged black hole discussed in \cite{Benini:2015eyy} which can be embedded into 11d supergravity in the regime where  $k$ is a small, fixed, number, $k\sim {\cal O}(1)$.  The precise embedding  equations are  given  \cite{Cvetic:1999xp}:
\bea
\label{Eq:11d}
ds_{11}^2 &=&\Delta^{2/3}ds_4^2+g^{-2}\Delta^{-1/3}\sum\limits_{i=1}^4 X^{-1}_i (d\mu^2_i +\mu_i^2 (d\phi_i+g A^{(i)})^2, \nonumber \\
F_4&=& 2g \sum\limits_{i=1}^4 \left(X_i^2\mu_i^2-\Delta X_i\right)\epsilon_{(4)}+\frac{1}{2g}\sum\limits_{i=1}^4 X_i^{-1}\bar{\star}\, dX_i \wedge d(\mu_i^2) \nonumber \\
&&-\frac{1}{2g^2}\sum\limits_{i=1}^4 X_i^{-2}d(\mu_i^2)\wedge (d\psi_i +g A^i)\wedge \bar{\star}F_{(2)}^i, \nonumber \\
\Delta&=& \sum\limits_{i=1}^4X_i \mu_i^2,
\eea
$ds_4^2$ denotes the four-dimensional metric, $X_i$ are scalar fields and $\sum\limits_{i=1}^4\mu_i^2=1$. 
For more details we refer the reader to \cite{Benini:2015eyy} and \cite{Cvetic:1999xp}. We will be very minimalistic and use the properties that we ultimately need.  In particular, the key ingredient is a reduction of this solution as to fit in the form of IIA background in Equation \ref{Eq:GravityBack}. The concrete technical problem we need to overcome is that the Ansatz above in Equation (\ref{Eq:11d}) is written for a parametrization of the $S^7$ which is not the the $U(1)$ bundle over $\mathbb{CP}^3$ used in Equation (\ref{Eq:GravityBack}).  Details of the appropriate coordinate changes are worked out in Appendix \ref{App:S7}. Our main concern is that the seven-dimensional part of the metric background takes the form 
\begin{equation}
    ds^2_7=\frac{4}{\Delta^{\frac{1}{3}}}\sum_{i=1}^4 \frac{1}{X_i}(d\mu_i^2+\mu_i^2(d\phi_i+\frac{n_i}{2}\cos\theta d\phi)^2),
\end{equation}
where we denote the coordinates on $S^2$ by $(\theta, \phi)$. Given the explicit form of the gauge fields  $A_i=(n_i/2)\cos\theta d\phi$ we understand how $\mathbb{CP}^3$ is deformed.  

Alternatively, and perhaps more conveniently, we might simply take the  four-dimensional point of view.  Most of the analysis of the solution was already performed in \cite{Benini:2015eyy} and rigorous approaches to holographic renormalization of this background were recently presented in \cite{Halmagyi:2017hmw,Cabo-Bizet:2017xdr,BenettiGenolini:2019jdz}. The bottom line is that given the prepotential the  4d ${\cal N}=2$ gauged supergravity theory, $F= -2i \sqrt{X^0X^1X^2X^3}$ the entropy of the background is simply obtained by 
\bea
S_{BH}&=& \frac{2\pi g^2}{G_{4D}}\sqrt{X_1(r_h)X_2(r_h)X_3(r_h)X_4(r_h)}.
\eea
At this point all that is required is that only aspect we need to modify in \cite{Benini:2015eyy} is the relation between the Newton's constant and the field theory values which can essentially be read off form the background in Equation (\ref{Eq:GravityBack}). In the 't Hooft limit we have
\be
\frac{2\, g^2}{G_{4D}}=\frac{2\sqrt{2}}{3}\,\, \frac{N^2}{\sqrt{\lambda}}.
\ee

We will not attempt to  compute the logarithmic in $\lambda$ corrections starting from one-loop supergravity in this work.  We, however, understand where this corrections might come from and will likely return to this problem in the future. 

\section{Conclusions}\label{Sec:Conclusions}

In this paper we have studied the leading and sub-leading structures of the topologically twisted index of ABJM theory in the 't Hooft limit. We have analytically obtained the leading $N^2/\sqrt{\lambda}$ behavior that matches precisely with the dual Bekenstein-Hawking  entropy of magnetically charged asymptotically AdS$_4$  black holes embedded in IIA string theory on AdS$_4\times \mathbb{CP}^3$. 

There was probably no doubts in the practitioner's mind that the 't Hooft limit will basically extend the  results of the M-theory limit, just as the free energy on $S^3$ explained the two supergravity scalings \cite{Drukker:2010nc}; this intuition is rightfully rooted in the idea of planar dominance on the field theory side.  In this paper we have completely clarified the mechanism through which that takes place by presenting various details of the corresponding eigenvalues; in particular, we have clarified  the scaling of the eigenvalues changes as we move from one limit to the other. 

Beyond the leading scaling with $N$, that is, the $N^{3/2}$ versus $N^2$ scaling, important salient differences between the large $N$, $k$--fixed limit and the 't Hooft limit show up at sub-leading orders.  The quantum expansion of the observables follow widely different patterns and our work here is one first step in the direction of exploring the topologically twisted index in the 't Hooft limit which plays a central role in microscopic counting of the black hole entropy on the dual IIA gravity side.  We have explored the topologically twisted index at genus one in the topological expansion. Our results on this part are mostly numerical but the structure constitute important predictions that should be reproduced on the gravitational type IIA string theory side. It would be quite interesting to explore the entropy of the corresponding black holes beyond the leading order and, in particular, to reproduce the $\log(\lambda)$ terms as a quantum one-loop computation in 10d type IIA supergravity.

The  topologically twisted index  of a certain Chern-Simons matter theory with $SU(N)$ gauge group at level $k$ has recently been matched with the corresponding black hole entropy in massive IIA gravity  \cite{Hosseini:2017fjo,Benini:2017oxt} elaborating on previous work  \cite{Guarino:2017eag,Guarino:2015jca,Varela:2015uca}. More recently, the topologically twisted index has been explored beyond the leading order in  $N$ and the coefficient of the logarithm in $N$ term has been determined \cite{Liu:2018bac}. It would be interesting to extend the analysis  in that case to cover the 't Hooft limit.   We have, however, indicated why a logarithmic in $\lambda$ term is very sub-leading in the topological expansion as it appears already as a sub-leading term in the $N^0$ part of the twisted partition function which is, itself,  sub-leading to the $N^2/\sqrt{\lambda}$ term. We have, nevertheless, presented some very explicit results in the special case where all the fugacities are the same, thus numerically confirming a similar structure in the genus-one term in the topological expansion of the free energy on $S^3$ and the topologically twisted index treated here.  

It would be interesting to discuss more generic field theories using the methods displayed in this manuscript and in \cite{Liu:2017vll,Liu:2018bac}.  It is tantalizing to surmise the existence of certain universality at the sub-leading order. For the topologically twisted index in  the large $N$, $k$--fixed limit, we hope to show that there is  certain universality in the coefficient of $\log N$ in  various classes of theories. The universality of the coefficient of the logarithm of $N$ for the free energy on $S^3$ for a large class of 3d  field theories was established in  \cite{Marino:2011eh,Fuji:2011km} as a consequence of properties of the Airy function. This field theoretic universality of the logarithmic in $N$ term has been reproduced on the gravity side in the M-theory limit in \cite{Bhattacharyya:2012ye} where it is the result of certain topological properties of seven-dimensional manifolds on which the dual  eleven-dimensional supergravity is compactified.  We anticipate that a similar phenomenon might be present for the topologically twisted indices based on direct numerical analysis and further by arguments of the supergravity description, we will report our findings in an upcoming publication. Armed with these results in the large $N$, $k$--fixed limit it seems likely that the 't Hooft limit will also lead to a certain universality in the sub-leading terms of the $1/N$ expansion. However, given the structure of the topological expansion the potential logarithmic term turns out to be logarithmic in  $\lambda$ and becomes technically harder to extract. 

On the more speculative side it would be quite interesting to explore whether the topological expansion of the index follows the free energy on $S^3$ in the sense that the coefficients of the topological expansion $\log Z=\sum g_s^{2g-2} F_g$ lead to $F_g$ with interesting modular properties. Such structure might not only clarify conceptually the nature of the expansion but will also provide an analytic understanding of the growth in the number of degrees of freedom setting up further high precision comparison with the gravitational side. 

\section*{Acknowledgments}
We are thankful to F. Benini, A. Faraggi, F. Ferrari, A. Gonz\'alez Lezcano, L. Griguolo, J. Hong, J. T.  Liu, J. F. Morales, K. Okuyama, P. Putrov, I. Yaakov and A. Zaffaroni for comments. This work is partially supported by the US Department of Energy under Grant No. DE-SC0007859.

\appendix

\section{Algorithmic details}\label{App:Algo}
Let us discuss various technical details regarding the numerical algorithms for solving the Bethe Ansatz Equations. A flow chart of the main algorithm is illustrated in table \ref{tbl:iteration}. We use $\{u^I_i,\tilde{u}^I_j\}$ to represent the starting point and $\{u^E_i,\tilde{u}^E_j\}$ to represent the exact solution of the variables $\{u_i,\tilde{u}_j\}$. The starting poing $\{u^I_i,\tilde{u}^I_j\}^{(o)}$ of the whole algorithm is the leading order eigenvalue distribution obtained in \cite{Benini:2015eyy} and the exact solution $\{u^E_i,\tilde{u}^E_j\}^{(o)}$ to the BAE with $k=1$ and small $N$ (for example, $N=50$) can be obtained using FindRoot in Mathematica as implemented in \cite{Liu:2017vll}. Because assigning a value to $k$ is equivalent to assigning a value to $\lambda \,(=N/k)$, we will use $k$ to illustrate the numerical algorithm in the following.

\begin{table}[h!]
\centering
\begin{tabular}{|c|c c c c c c|}
\hline
$\{u_i,\tilde{u}_j\}$ & $N_o$ & $\cdots$ & $N_n$ & $N_{n+1}$ & $\cdots$ & $N_f$ \\
\hline
$k_o$ & \multicolumn{1}{l}{$\{u_i,\tilde{u}_j\}^{(o)}$} & $\rightarrow$ & \multicolumn{1}{l}{$\{u_i,\tilde{u}_j\}^{(n)}$} & \multicolumn{1}{l}{$\{u_i,\tilde{u}_j\}^{(n+1)}$} & $\rightarrow$ & \\
$\vdots$ & $\downarrow$ & & & & & \\
$k_m$ & \multicolumn{1}{l}{$\{u_i,\tilde{u}_j\}^{(m)}$} & & & & & \\
$k_{m+1}$ & \multicolumn{1}{l}{$\{u_i,\tilde{u}_j\}^{(m+1)}$} & & & & & \\
$\vdots$ & $\downarrow$ & & & & & \\
$k_f$ & & & & & & $\{u_i,\tilde{u}_j\}^{(f)}$ \\
\hline
\end{tabular}
\caption{The flow chart of the iterative algorithm for solving the BAE}
\label{tbl:iteration}
\end{table}

The iterative algorithm contains two parts. The first part is for the BAE with a fixed $N$ but different values of $k$, namely the vertical direction. The solutions in the same column have the same dimension $2N$. The behavior of the eigenvalues in Equation (\ref{Eq:leadingk}) implies an iterative relation 
\bea
\{u^I_i,\tilde{u}^I_j\}^{(m+1)}&=&i\sqrt{\frac{k_m}{k_{m+1}}}{\rm Im}\Big(\{u^E_i,\tilde{u}^E_j\}^{(m)}\Big)+{\rm Re}\Big(\{u^E_i,\tilde{u}^E_j\}^{(m)}\Big)-\frac{\pi}{k_m}+\frac{\pi}{k_{m+1}}, \nonumber \\
\{u^E_i,\tilde{u}^E_j\}^{(m+1)}&=&{\rm FindRoot}\Big[{\rm BAE}(N_o,k_{m+1}), \{u^I_i,\tilde{u}^I_j\}^{(m+1)}\Big].
\eea

Thus the first iterative algorithm is $k$ iterative algorithm (Algorithm \ref{algo:kalgo}). It is worth noting that the number of the step $N_{ks}$ for iterating over $k$ can be set to 1 for the special case, but has to satisfy the condition $k_s=(k_f/k_o)^{1/N_{ks}} \gtrapprox 1$ for the general cases.
\vspace{12pt}

\begin{algorithm}[H]
\KwIn{$N_o,k_o,\{u^E_i,\tilde{u}^E_j\}^{(o)},k_f,N_{ks}$\;}
Initialize $\{u^E_i,\tilde{u}^E_j\}=\{u^E_i,\tilde{u}^E_j\}^{(o)};k_s=(k_f/k_o)^{1/N_{ks}}$;\\
\For {$k=k_o\times k_s,k_o\times k_s^2,\cdots, $ \KwTo $k=k_f$}{
$\{u^I_i,\tilde{u}^I_j\}:=i\sqrt{1/k_s}{\rm Im}\Big(\{u^E_i,\tilde{u}^E_j\}\Big)+{\rm Re}\Big(\{u^E_i,\tilde{u}^E_j\}\Big)-\pi/(k/k_s)+\pi/k$\;
$\{u^E_i,\tilde{u}^E_j\}:={\rm FindRoot}\Big[{\rm BAE}(N_o,k),\{u^I_i,\tilde{u}^I_j\}\Big]$\;
}
\caption{$k$ iterative algorithm}
\label{algo:kalgo}
\end{algorithm}

The second part is for the BAE with a fixed $k$ but different values of $N$, namely the horizontal direction. Because the dimension of the solutions in the same row is dependent on $N$, the method of interpolation should be used in this iterative algorithm. Also implied by the behavior form in Equation (\ref{Eq:leadingk}), the iterative relation is
\bea
t^{(n)}(i)&=&{\rm Interpolation}\bigg[\bigg\{\frac{i-1}{N_n-1},\sqrt{\frac{k_o}{N_n}}{\rm Im}\Big((u^E_i)^{(n)}\Big)\bigg\},i=1,2,\cdots,N_n\bigg], \nonumber \\
dv^{(n)}(i)&=&{\rm Interpolation}\bigg[\bigg\{\frac{i-1}{N_n-1},-2\bigg({\rm Re}\Big((u^E_i)^{(n)}\Big)-\frac{\pi}{k_o}\bigg)\bigg\},i=1,2,\cdots,N_n\bigg], \nonumber \\
(u^I_i)^{(n+1)}&=&i\sqrt{\frac{N_{n+1}}{k_o}}t^{(n)}\bigg(\frac{i-1}{N_{n+1}-1}\bigg)+\frac{\pi}{k_o}-\frac{1}{2}dv^{(n)}\bigg(\frac{i-1}{N_{n+1}-1}\bigg), i=1,2,\cdots,N_{n+1}, \nonumber \\
(\tilde{u}^I_j)^{(n+1)}&=&i\sqrt{\frac{N_{n+1}}{k_o}}t^{(n)}\bigg(\frac{j-1}{N_{n+1}-1}\bigg)+\frac{\pi}{k_o}+\frac{1}{2}dv^{(n)}\bigg(\frac{j-1}{N_{n+1}-1}\bigg), j=1,2,\cdots,N_{n+1}, \nonumber \\
\{u^E_i,\tilde{u}^E_j\}^{(n+1)}&=&{\rm FindRoot}\Big[{\rm BAE}(N_{n+1},k_o), \{u^I_i,\tilde{u}^I_j\}^{(n+1)}\Big].
\eea

Thus the second iterative algorithm is $N$ iterative algorithm (Algorithm \ref{algo:nalgo}). It is worth noting that the step $N_{s}$ for iterating over $N$ is a natural number and much less than $N(N\sim100)$. In the line $(*)$ of the algorithm, $k:=k_o$ is for the M-theory limit. Since $N$ changes smoothly, we can set $k:=(N/N_o)k_o$ for the 't Hooft limit. 

\begin{algorithm}[H]
\KwIn{$N_o,k_o,\{u^E_i,\tilde{u}^E_j\}^{(o)},N_f,N_s$;}
Initialize $\{u^E_i,\tilde{u}^E_j\}=\{u^E_i,\tilde{u}^E_j\}^{(o)};N_I=N_o;k_I=k_o$\;
\For {$N=N_o+N_s,N_o+2N_s,\cdots, $ \KwTo $N=N_f$}{
$t(i):={\rm Interpolation}\Big[\Big\{\frac{i-1}{N_I-1},\sqrt{\frac{k_I}{N_I}}{\rm Im}\Big(u^E_i\Big)\Big\},i=1,2,\cdots,N_I\Big]$\;
$dv(i):={\rm Interpolation}\Big[\Big\{\frac{i-1}{N_I-1},-2\Big({\rm Re}\Big(u^E_i\Big)-\frac{\pi}{k_I}\Big)\Big\},i=1,2,\cdots,N_I\Big]$\;
$k:=k_o$; \tcp{$(*) \ k:=(N/N_o)k_o;$}
\For {$i=1:N$}{
$u^I_i:=i\sqrt{\frac{N}{k}}t\Big(\frac{i-1}{N-1}\Big)+\frac{\pi}{k}-\frac{1}{2}dv\Big(\frac{i-1}{N-1}\Big)$\;
$\tilde{u}^I_i:=i\sqrt{\frac{N}{k}}t\Big(\frac{i-1}{N-1}\Big)+\frac{\pi}{k}+\frac{1}{2}dv\Big(\frac{i-1}{N-1}\Big)$\;
}
$N_I:=N;k_I:=k$\;
$\{u^E_i,\tilde{u}^E_j\}:={\rm FindRoot}\Big[{\rm BAE}(N,k),\{u^I_i,\tilde{u}^I_j\}\Big]$\;
}
\caption{$N$ iterative algorithm}
\label{algo:nalgo}
\end{algorithm}

Theoretically, the BAE with any values of $N$ and $k$  can be solved using the combination of the $k$ iterative algorithm and $N$ iterative algorithm. However, with increasing $N$, the scale and complexity of the BAE increases so that it  takes much more time to arrive at the final solution, especially for the cases with general fugacities. Therefore, we choose to obtain the numerical solutions up to $N=300\sim400$. As an example, the detailed information of the numerical computation in Table \ref{tbl:FitlogZDat} is shown in Table \ref{tbl:NumInfo}. WP means WorkingPrecision in Mathematica, BAE time means the time of solving the BAE and ${\rm Re}\log Z$ time means the time of calculating the index.

\begin{table}[h]
\centering
\begin{tabular}{|c|c|c|c|c|c|c|c|c|c|}
\hline
$\Delta_a$ & $\lambda$ & $k_o$ & $N_o$ & $N$ range & $N_{ks}$ & $N_s$ & WP & BAE time & ${\rm Re}\log Z$ time \\
\hline
Special & $1,5,10$ & $1$ & $100$ & $100\sim300$ & $1$ & $10$ & $200$ & $2$ hrs & $40$ min \\
\hline
\multirow{3}{*}{General} & $1$ & $1$ & $50$ & $50\sim300$ & $10$ & $5$ & $600$ & $33$ hrs & $3$ hrs $40$ min \\
\cline{2-10}
& $5$ & $1$ & $50$ & $50\sim300$ & $10$ & $10$ & $300$ & $10$ hrs & $3$ hrs $22$ min \\
\cline{2-10}
& $10$ & $1$ & $50$ & $50\sim300$ & $10$ & $10$ & $300$ & $10$ hrs & $3$ hrs $30$ min \\
\hline
\end{tabular}
\caption{The information of the numerical computation in Table \ref{tbl:FitlogZDat}.}
\label{tbl:NumInfo}
\end{table}

\section{Parametrizations of $S^7$}\label{App:S7}

In this appendix we describe the embedding of magnetically charged asymptotically AdS$_4$ black holes, such as those of  \cite{Cacciatori:2009iz} in eleven and ten-dimensional contexts.  We first discuss the setup of  \cite{Cvetic:1999xp} paying particular attention to the details of the reduction to 10d. 
Let us write $S^7$ as a $U(1)$ bundle over $\mathbb{CP}^3$
\bea
ds^2_{S^7}&=& \frac{1}{16} (d\zeta +A)^2\nonumber \\
&+& \frac{1}{4}\bigg[d\alpha^2+\cos^2\frac{\a}{2}(d\th_1^2+\sin^2\th_1^2 d\varphi_1^2) + \sin^2\frac{\a}{2}(d\th_2^2+\sin^2\th_2^2 d\varphi_2^2) \nonumber \\
&+& \sin^2\frac{\a}{2}\cos^2\frac{\a}{2}(d\chi + \cos\th_1 d\varphi_1-\cos\th_2 d\varphi_2)\bigg],
\eea
where 
$$
A= \cos\a d\chi +2\cos^2\frac{\a}{2}\cos\th_1 d\varphi_1+ 2\sin^2\frac{\a}{2}\cos\th_2 d\varphi_2.
$$
The above metric can be reached by considering \cite{Cvetic:2000yp}  
\bea
z_1&=& \cos\frac{\a}{2}\cos\frac{\th_1}{2}\exp\big[ i \left(2\varphi_1+\chi+\zeta\right)/4\big], \nonumber \\
z_2&=& \cos\frac{\a}{2}\sin\frac{\th_1}{2}\exp\big[ i \left(-2\varphi_1+\chi+\zeta\right)/4\big], \nonumber \\
z_3&=& \cos\frac{\a}{2}\cos\frac{\th_2}{2}\exp\big[ i \left(2\varphi_2-\chi+\zeta\right)/4\big], \nonumber \\
z_4&=& \cos\frac{\a}{2}\sin\frac{\th_2}{2}\exp\big[ i \left(-2\varphi_2-\chi+\zeta\right)/4\big],
\eea
which satisfy $\sum\limits_{i=1}^4|z_i|^2=1$.

We have different options for writing $S^7$ as foliations of spheres, 
\be
ds^2_{S^7}=\sum\limits_{i=1}^4(d\mu_i^2 +\mu_i^2d\phi_i^2).
\ee

For example, taking
\bea
\mu_1&=&\sin\th, \nonumber \\
\mu_2&=& \cos\th\sin\varphi, \nonumber \\
\mu_2&=&\cos\th\cos\varphi \sin \psi, \nonumber\\
\mu_4&=&\cos\th\cos\varphi\cos\psi
\eea
Leads to 
\bea
ds^2_{S^7}&=& d\th^2 +\sin^2\th d\phi_1^2\nonumber \\
&+& \cos^2 \th \left(d\varphi^2 +\cos^2 \varphi (d\psi^2 +\sin^2\psi d\phi_3^2 +\cos^2\psi d\phi_4^2) +\sin^2\varphi d\phi_2^2\right),
\eea
which is a foliation of $S^5\times S^1$. For a foliation of $S^7$ over $S^3\times S^3$ we need:
\bea
\mu_1&=&  \cos\frac{\a}{2}\cos\frac{\th_1}{2} \nonumber \\
\mu_2&=& \sin\frac{\a}{2}\sin\frac{\th_1}{2} ,\nonumber \\
\mu_2&=&  \cos\frac{\a}{2}\cos\frac{\th_2}{2}, \nonumber\\
\mu_4&=&  \sin\frac{\a}{2}\sin\frac{\th_2}{2},
\eea
which leads to 
\bea
ds^2_{S^7}&=& d\a^2 + \cos^2\a\left(d\th_1^2 +\cos^2\th_1 d\phi_1^2 +\sin^2\th_1 d\phi_2^2\right) \nonumber\\
&+&  \sin^2\a\left(d\th_2^2 +\cos^2\th_2 d\phi_3^2 +\sin^2\th_2 d\phi_4^2\right).
\eea
It  becomes clear in either of these parametrizations that  the fiber in $\zeta$ is obtained as the sum of the angles $\phi_i$ in the $(\mu_i, \phi_i)$ parametrization. 
\be
\zeta=\frac{1}{4}(\phi_1+\phi_2+\phi_3+\phi_4).
\ee
This is the angle for which, uppon reduction to ten dimensions one recovers the background in Equation (\ref{Eq:GravityBack}) in the absences of charges and for trivial scalar fields. 

\bibliographystyle{JHEP}
\bibliography{BHLocalization}

\providecommand{\href}[2]{#2}\begingroup\raggedright\begin{thebibliography}{10}

\bibitem{Aharony:2008ug}
O.~Aharony, O.~Bergman, D.~L. Jafferis and J.~Maldacena, \emph{{$\mathcal{N}=6$
  superconformal Chern-Simons-matter theories, M2-branes and their gravity
  duals}}, \href{https://doi.org/10.1088/1126-6708/2008/10/091}{\emph{JHEP}
  {\bfseries 10} (2008) 091},
  [\href{https://arxiv.org/abs/0806.1218}{{\ttfamily 0806.1218}}].

\bibitem{Drukker:2010nc}
N.~Drukker, M.~Marino and P.~Putrov, \emph{{From weak to strong coupling in
  ABJM theory}},
  \href{https://doi.org/10.1007/s00220-011-1253-6}{\emph{Commun.Math.Phys.}
  {\bfseries 306} (2011) 511--563},
  [\href{https://arxiv.org/abs/1007.3837}{{\ttfamily 1007.3837}}].

\bibitem{Drukker:2011zy}
N.~Drukker, M.~Marino and P.~Putrov, \emph{{Nonperturbative aspects of ABJM
  theory}}, \href{https://doi.org/10.1007/JHEP11(2011)141}{\emph{JHEP}
  {\bfseries 11} (2011) 141},
  [\href{https://arxiv.org/abs/1103.4844}{{\ttfamily 1103.4844}}].

\bibitem{Benini:2015noa}
F.~Benini and A.~Zaffaroni, \emph{{A topologically twisted index for
  three-dimensional supersymmetric theories}},
  \href{https://doi.org/10.1007/JHEP07(2015)127}{\emph{JHEP} {\bfseries 07}
  (2015) 127}, [\href{https://arxiv.org/abs/1504.03698}{{\ttfamily
  1504.03698}}].

\bibitem{Honda:2015yha}
M.~Honda and Y.~Yoshida, \emph{{Supersymmetric index on $T^2 \times S^2$ and
  elliptic genus}},  \href{https://arxiv.org/abs/1504.04355}{{\ttfamily
  1504.04355}}.

\bibitem{Closset:2015rna}
C.~Closset, S.~Cremonesi and D.~S. Park, \emph{{The equivariant A-twist and
  gauged linear sigma models on the two-sphere}},
  \href{https://doi.org/10.1007/JHEP06(2015)076}{\emph{JHEP} {\bfseries 06}
  (2015) 076}, [\href{https://arxiv.org/abs/1504.06308}{{\ttfamily
  1504.06308}}].

\bibitem{Benini:2016hjo}
F.~Benini and A.~Zaffaroni, \emph{{Supersymmetric partition functions on
  Riemann surfaces}},  \href{https://arxiv.org/abs/1605.06120}{{\ttfamily
  1605.06120}}.

\bibitem{Closset:2016arn}
C.~Closset and H.~Kim, \emph{{Comments on twisted indices in 3d supersymmetric
  gauge theories}},  \href{https://arxiv.org/abs/1605.06531}{{\ttfamily
  1605.06531}}.

\bibitem{Closset:2017zgf}
C.~Closset, H.~Kim and B.~Willett, \emph{{Supersymmetric partition functions
  and the three-dimensional A-twist}},
  \href{https://doi.org/10.1007/JHEP03(2017)074}{\emph{JHEP} {\bfseries 03}
  (2017) 74}, [\href{https://arxiv.org/abs/1701.03171}{{\ttfamily
  1701.03171}}].

\bibitem{Closset:2018ghr}
C.~Closset, H.~Kim and B.~Willett, \emph{{Seifert fibering operators in 3d
  $\mathcal{N}=2$ theories}},
  \href{https://arxiv.org/abs/1807.02328}{{\ttfamily 1807.02328}}.

\bibitem{Benini:2015eyy}
F.~Benini, K.~Hristov and A.~Zaffaroni, \emph{{Black hole microstates in
  AdS$_{4}$ from supersymmetric localization}},
  \href{https://doi.org/10.1007/JHEP05(2016)054}{\emph{JHEP} {\bfseries 05}
  (2016) 054}, [\href{https://arxiv.org/abs/1511.04085}{{\ttfamily
  1511.04085}}].

\bibitem{Benini:2016rke}
F.~Benini, K.~Hristov and A.~Zaffaroni, \emph{{Exact microstate counting for
  dyonic black holes in AdS$_4$}},
  \href{https://doi.org/10.1016/j.physletb.2017.05.076}{\emph{Phys. Lett.}
  {\bfseries B771} (2017) 462--466},
  [\href{https://arxiv.org/abs/1608.07294}{{\ttfamily 1608.07294}}].

\bibitem{Cabo-Bizet:2017jsl}
A.~Cabo-Bizet, V.~I. Giraldo-Rivera and L.~A. Pando~Zayas, \emph{{Microstate
  counting of AdS$_{4}$ hyperbolic black hole entropy via the topologically
  twisted index}}, \href{https://doi.org/10.1007/JHEP08(2017)023}{\emph{JHEP}
  {\bfseries 08} (2017) 023},
  [\href{https://arxiv.org/abs/1701.07893}{{\ttfamily 1701.07893}}].

\bibitem{Azzurli:2017kxo}
F.~Azzurli, N.~Bobev, P.~M. Crichigno, V.~S. Min and A.~Zaffaroni, \emph{{A
  universal counting of black hole microstates in AdS$_{4}$}},
  \href{https://doi.org/10.1007/JHEP02(2018)054}{\emph{JHEP} {\bfseries 02}
  (2018) 054}, [\href{https://arxiv.org/abs/1707.04257}{{\ttfamily
  1707.04257}}].

\bibitem{Hosseini:2017fjo}
S.~M. Hosseini, K.~Hristov and A.~Passias, \emph{{Holographic microstate
  counting for AdS$_{4}$ black holes in massive IIA supergravity}},
  \href{https://doi.org/10.1007/JHEP10(2017)190}{\emph{JHEP} {\bfseries 10}
  (2017) 190}, [\href{https://arxiv.org/abs/1707.06884}{{\ttfamily
  1707.06884}}].

\bibitem{Benini:2017oxt}
F.~Benini, H.~Khachatryan and P.~Milan, \emph{{Black hole entropy in massive
  Type IIA}}, \href{https://doi.org/10.1088/1361-6382/aa9f5b}{\emph{Class.
  Quant. Grav.} {\bfseries 35} (2018) 035004},
  [\href{https://arxiv.org/abs/1707.06886}{{\ttfamily 1707.06886}}].

\bibitem{Hosseini:2018qsx}
S.~M. Hosseini, \emph{{Black hole microstates and supersymmetric
  localization}}, Ph.D. thesis, Milan Bicocca U., 2018-02.
\newblock \href{https://arxiv.org/abs/1803.01863}{{\ttfamily 1803.01863}}.

\bibitem{Zaffaroni:2019dhb}
A.~Zaffaroni, \emph{{Lectures on AdS Black Holes, Holography and
  Localization}},  2019, \href{https://arxiv.org/abs/1902.07176}{{\ttfamily
  1902.07176}}.

\bibitem{Liu:2017vll}
J.~T. Liu, L.~A. Pando~Zayas, V.~Rathee and W.~Zhao, \emph{{Toward Microstate
  Counting Beyond Large N in Localization and the Dual One-loop Quantum
  Supergravity}}, \href{https://doi.org/10.1007/JHEP01(2018)026}{\emph{JHEP}
  {\bfseries 01} (2018) 026},
  [\href{https://arxiv.org/abs/1707.04197}{{\ttfamily 1707.04197}}].

\bibitem{Liu:2018bac}
J.~T. Liu, L.~A. Pando~Zayas and S.~Zhou, \emph{{Subleading Microstate Counting
  in the Dual to Massive Type IIA}},
  \href{https://arxiv.org/abs/1808.10445}{{\ttfamily 1808.10445}}.

\bibitem{Liu:2017vbl}
J.~T. Liu, L.~A. Pando~Zayas, V.~Rathee and W.~Zhao, \emph{{One-Loop Test of
  Quantum Black Holes in anti–de Sitter Space}},
  \href{https://doi.org/10.1103/PhysRevLett.120.221602}{\emph{Phys. Rev. Lett.}
  {\bfseries 120} (2018) 221602},
  [\href{https://arxiv.org/abs/1711.01076}{{\ttfamily 1711.01076}}].

\bibitem{Jeon:2017aif}
I.~Jeon and S.~Lal, \emph{{Logarithmic Corrections to Entropy of Magnetically
  Charged AdS$_{4}$ Black Holes}},
  \href{https://doi.org/10.1016/j.physletb.2017.09.026}{\emph{Phys. Lett.}
  {\bfseries B774} (2017) 41--45},
  [\href{https://arxiv.org/abs/1707.04208}{{\ttfamily 1707.04208}}].

\bibitem{Hristov:2018lod}
K.~Hristov, I.~Lodato and V.~Reys, \emph{{On the quantum entropy function in 4d
  gauged supergravity}},
  \href{https://doi.org/10.1007/JHEP07(2018)072}{\emph{JHEP} {\bfseries 07}
  (2018) 072}, [\href{https://arxiv.org/abs/1803.05920}{{\ttfamily
  1803.05920}}].

\bibitem{Gang:2018hjd}
D.~Gang and N.~Kim, \emph{{Large $N$ twisted partition functions in 3d-3d
  correspondence and Holography}},
  \href{https://doi.org/10.1103/PhysRevD.99.021901}{\emph{Phys. Rev.}
  {\bfseries D99} (2019) 021901},
  [\href{https://arxiv.org/abs/1808.02797}{{\ttfamily 1808.02797}}].

\bibitem{Gang:2019uay}
D.~Gang, N.~Kim and L.~A. Pando~Zayas, \emph{{Precision Microstate Counting for
  the Entropy of Wrapped M5-branes}},
  \href{https://arxiv.org/abs/1905.01559}{{\ttfamily 1905.01559}}.

\bibitem{Herzog:2010hf}
C.~P. Herzog, I.~R. Klebanov, S.~S. Pufu and T.~Tesileanu, \emph{{Multi-Matrix
  Models and Tri-Sasaki Einstein Spaces}},
  \href{https://doi.org/10.1103/PhysRevD.83.046001}{\emph{Phys. Rev.}
  {\bfseries D83} (2011) 046001},
  [\href{https://arxiv.org/abs/1011.5487}{{\ttfamily 1011.5487}}].

\bibitem{Hosseini:2016tor}
S.~M. Hosseini and A.~Zaffaroni, \emph{{Large $N$ matrix models for 3d ${\cal
  N}=2$ theories: twisted index, free energy and black holes}},
  \href{https://doi.org/10.1007/JHEP08(2016)064}{\emph{JHEP} {\bfseries 08}
  (2016) 064}, [\href{https://arxiv.org/abs/1604.03122}{{\ttfamily
  1604.03122}}].

\bibitem{Hosseini:2016ume}
S.~M. Hosseini and N.~Mekareeya, \emph{{Large $N$ topologically twisted index:
  necklace quivers, dualities, and Sasaki-Einstein spaces}},
  \href{https://doi.org/10.1007/JHEP08(2016)089}{\emph{JHEP} {\bfseries 08}
  (2016) 089}, [\href{https://arxiv.org/abs/1604.03397}{{\ttfamily
  1604.03397}}].

\bibitem{Marino:2009jd}
M.~Marino and P.~Putrov, \emph{{Exact Results in ABJM Theory from Topological
  Strings}}, \href{https://doi.org/10.1007/JHEP06(2010)011}{\emph{JHEP}
  {\bfseries 06} (2010) 011},
  [\href{https://arxiv.org/abs/0912.3074}{{\ttfamily 0912.3074}}].

\bibitem{Ooguri:2002gx}
H.~Ooguri and C.~Vafa, \emph{{World sheet derivation of a large N duality}},
  \href{https://doi.org/10.1016/S0550-3213(02)00620-X}{\emph{Nucl. Phys.}
  {\bfseries B641} (2002) 3--34},
  [\href{https://arxiv.org/abs/hep-th/0205297}{{\ttfamily hep-th/0205297}}].

\bibitem{Okuyama:2016deu}
K.~Okuyama, \emph{{Instanton Corrections of $1/6$ BPS Wilson Loops in ABJM
  Theory}}, \href{https://doi.org/10.1007/JHEP09(2016)125}{\emph{JHEP}
  {\bfseries 09} (2016) 125},
  [\href{https://arxiv.org/abs/1607.06157}{{\ttfamily 1607.06157}}].

\bibitem{Liu:2010gz}
J.~T. Liu and R.~Minasian, \emph{{Computing $1/N^2$ corrections in AdS/CFT}},
  \href{https://arxiv.org/abs/1010.6074}{{\ttfamily 1010.6074}}.

\bibitem{Sen:2011ba}
A.~Sen, \emph{{Logarithmic Corrections to $\mathcal{N}=2$ Black Hole Entropy:
  An Infrared Window into the Microstates}},
  \href{https://doi.org/10.1007/s10714-012-1336-5}{\emph{Gen. Rel. Grav.}
  {\bfseries 44} (2012) 1207--1266},
  [\href{https://arxiv.org/abs/1108.3842}{{\ttfamily 1108.3842}}].

\bibitem{Sen:2012cj}
A.~Sen, \emph{{Logarithmic Corrections to Rotating Extremal Black Hole Entropy
  in Four and Five Dimensions}},
  \href{https://doi.org/10.1007/s10714-012-1373-0}{\emph{Gen. Rel. Grav.}
  {\bfseries 44} (2012) 1947--1991},
  [\href{https://arxiv.org/abs/1109.3706}{{\ttfamily 1109.3706}}].

\bibitem{Sen:2014aja}
A.~Sen, \emph{{Microscopic and Macroscopic Entropy of Extremal Black Holes in
  String Theory}}, \href{https://doi.org/10.1007/s10714-014-1711-5}{\emph{Gen.
  Rel. Grav.} {\bfseries 46} (2014) 1711},
  [\href{https://arxiv.org/abs/1402.0109}{{\ttfamily 1402.0109}}].

\bibitem{Bhattacharyya:2012ye}
S.~Bhattacharyya, A.~Grassi, M.~Marino and A.~Sen, \emph{{A One-Loop Test of
  Quantum Supergravity}},
  \href{https://doi.org/10.1088/0264-9381/31/1/015012}{\emph{Class. Quant.
  Grav.} {\bfseries 31} (2014) 015012},
  [\href{https://arxiv.org/abs/1210.6057}{{\ttfamily 1210.6057}}].

\bibitem{Cvetic:1999xp}
M.~Cvetic, M.~J. Duff, P.~Hoxha, J.~T. Liu, H.~Lu, J.~X. Lu et~al.,
  \emph{{Embedding AdS black holes in ten-dimensions and eleven-dimensions}},
  \href{https://doi.org/10.1016/S0550-3213(99)00419-8}{\emph{Nucl. Phys.}
  {\bfseries B558} (1999) 96--126},
  [\href{https://arxiv.org/abs/hep-th/9903214}{{\ttfamily hep-th/9903214}}].

\bibitem{Halmagyi:2017hmw}
N.~Halmagyi and S.~Lal, \emph{{On the on-shell: the action of AdS$_{4}$ black
  holes}}, \href{https://doi.org/10.1007/JHEP03(2018)146}{\emph{JHEP}
  {\bfseries 03} (2018) 146},
  [\href{https://arxiv.org/abs/1710.09580}{{\ttfamily 1710.09580}}].

\bibitem{Cabo-Bizet:2017xdr}
A.~Cabo-Bizet, U.~Kol, L.~A. Pando~Zayas, I.~Papadimitriou and V.~Rathee,
  \emph{{Entropy functional and the holographic attractor mechanism}},
  \href{https://doi.org/10.1007/JHEP05(2018)155}{\emph{JHEP} {\bfseries 05}
  (2018) 155}, [\href{https://arxiv.org/abs/1712.01849}{{\ttfamily
  1712.01849}}].

\bibitem{BenettiGenolini:2019jdz}
P.~Benetti~Genolini, J.~M. Pérez~Ipiña and J.~Sparks, \emph{{Localization of
  the action in AdS/CFT}},  \href{https://arxiv.org/abs/1906.11249}{{\ttfamily
  1906.11249}}.

\bibitem{Guarino:2017eag}
A.~Guarino and J.~Tarrio, \emph{{BPS black holes from massive IIA on S$^{6}$}},
  \href{https://doi.org/10.1007/JHEP09(2017)141}{\emph{JHEP} {\bfseries 09}
  (2017) 141}, [\href{https://arxiv.org/abs/1703.10833}{{\ttfamily
  1703.10833}}].

\bibitem{Guarino:2015jca}
A.~Guarino, D.~L. Jafferis and O.~Varela, \emph{{String Theory Origin of Dyonic
  $ \mathcal{N}=8 $ Supergravity and Its Chern-Simons Duals}},
  \href{https://doi.org/10.1103/PhysRevLett.115.091601}{\emph{Phys. Rev. Lett.}
  {\bfseries 115} (2015) 091601},
  [\href{https://arxiv.org/abs/1504.08009}{{\ttfamily 1504.08009}}].

\bibitem{Varela:2015uca}
O.~Varela, \emph{{AdS$_{4}$ solutions of massive IIA from dyonic ISO(7)
  supergravity}}, \href{https://doi.org/10.1007/JHEP03(2016)071}{\emph{JHEP}
  {\bfseries 03} (2016) 071},
  [\href{https://arxiv.org/abs/1509.07117}{{\ttfamily 1509.07117}}].

\bibitem{Marino:2011eh}
M.~Marino and P.~Putrov, \emph{{ABJM theory as a Fermi gas}},
  \href{https://doi.org/10.1088/1742-5468/2012/03/P03001}{\emph{J. Stat. Mech.}
  {\bfseries 1203} (2012) P03001},
  [\href{https://arxiv.org/abs/1110.4066}{{\ttfamily 1110.4066}}].

\bibitem{Fuji:2011km}
H.~Fuji, S.~Hirano and S.~Moriyama, \emph{{Summing Up All Genus Free Energy of
  ABJM Matrix Model}},
  \href{https://doi.org/10.1007/JHEP08(2011)001}{\emph{JHEP} {\bfseries 08}
  (2011) 001}, [\href{https://arxiv.org/abs/1106.4631}{{\ttfamily 1106.4631}}].

\bibitem{Cacciatori:2009iz}
S.~L. Cacciatori and D.~Klemm, \emph{{Supersymmetric AdS$_{4}$ black holes and
  attractors}}, \href{https://doi.org/10.1007/JHEP01(2010)085}{\emph{JHEP}
  {\bfseries 01} (2010) 085},
  [\href{https://arxiv.org/abs/0911.4926}{{\ttfamily 0911.4926}}].

\bibitem{Cvetic:2000yp}
M.~Cvetic, H.~Lu and C.~N. Pope, \emph{{Consistent warped space Kaluza-Klein
  reductions, half maximal gauged supergravities and $CP^n$ constructions}},
  \href{https://doi.org/10.1016/S0550-3213(00)00708-2}{\emph{Nucl. Phys.}
  {\bfseries B597} (2001) 172--196},
  [\href{https://arxiv.org/abs/hep-th/0007109}{{\ttfamily hep-th/0007109}}].

\end{thebibliography}\endgroup

\end{document}